\documentclass[12pt]{article}

\pdfoutput=1

\usepackage{color}

\usepackage{amsmath,amssymb,latexsym}

\usepackage{cite}

\usepackage[T1]{fontenc}
\usepackage{times}

\usepackage{epstopdf}

\usepackage{graphics,epsfig}

\usepackage{caption}
\usepackage{fdsymbol}

\let\a=\alpha \let\b=\beta \let\g=\gamma \let\d=\delta \let\e=\epsilon
\let\i=\iota \let\k=\kappa
\let\l=\lambda \let\m=\mu \let\n=\nu \let\x=\xi \let\p=\pi 
\let\s=\sigma 
 \let\f=\phi  \let\y=\psi
\let\vp=\varphi 
\let\w=\omega       \let\D=\Delta \let\Th=\Theta \let\L=\Lambda
\let\X=\Xi  \let\S=\Sigma  \let\Y=\Psi
 
\let\la=\label  
  
 \def\bd{\begin{document}} \def\ed{\end{document}}
\def\ds{\documentstyle} \let\fr=\frac \let\bl=\bigl \let\br=\bigr
\let\Br=\Bigr \let\Bl=\Bigl
\let\bm=\bibitem
\let\na=\nabla
\def\tU{{\widetilde U}}
\let\pa=\partial \let\ov=\overline
\def\ie{{\it i.e.\ }}
\newcommand{\be}{\begin{equation}}
\newcommand{\ee}{\end{equation}}
\def\ba{\begin{array}}
\def\ea{\end{array}}

\def\bei{\begin{itemize}}
\def\eei{\end{itemize}}

\def\ben{\begin{enumerate}}
\def\een{\end{enumerate}}

\def\ft#1#2{{\textstyle{{\scriptstyle #1}\over {\scriptstyle #2}}}}
\def\fft#1#2{{#1 \over #2}}
\def\F#1#2{{ F_{#1}^{(#2)} }}
\def\cF#1#2{{ {\cal F}_{#1}^{(#2)} }}

\def\R{{\bf R}}
\def\sst#1{{\scriptscriptstyle #1}}
\def\oneone{\rlap 1\mkern4mu{\rm l}}
\def\e7{E_{7(+7)}}
\def\td{\tilde}
\def\wtd{\widetilde}
\def\im{{\rm i}}
\def\bog{Bogomol'nyi\ }
\newcommand{\ho}[1]{$\, ^{#1}$}
\newcommand{\hoch}[1]{$\, ^{#1}$}
\newcommand{\bea}{\begin{eqnarray}}
\newcommand{\eea}{\end{eqnarray}}
\newcommand{\ra}{\rightarrow}
\newcommand{\lra}{\longrightarrow}
\newcommand{\Lra}{\Leftrightarrow}
\newcommand{\ap}{\alpha^\prime}
\newcommand{\bp}{\tilde \beta^\prime}
\newcommand{\cB}{{\cal B}}
\newcommand{\cO}{{\cal O}}
\newcommand{\vecx}{\vec{x}}
\newcommand{\vecy}{\vec{y}}
\newcommand{\vecp}{\vec{p}}
\newcommand{\vecq}{\vec{q}}
\newcommand{\tr}{{\rm tr} }
\newcommand{\Tr}{{\rm Tr} }
\newcommand{\NP}{Nucl. Phys. }

\newcommand{\cL}{{\cal L}}
\newcommand{\cA}{{\cal A}}
\newcommand{\cT}{{\cal T}}
\newcommand{\cD}{{\cal D}}
\newcommand{\cH}{{\cal H}}

\def\th{\theta}

\def\sst#1{{\scriptscriptstyle #1}}
\def\0{{\sst{(0)}}}
\def\1{{\sst{(1)}}}
\def\2{{\sst{(2)}}}
\def\3{{\sst{(3)}}}
\def\4{{\sst{(4)}}}
\def\5{{\sst{(5)}}}
\def\6{{\sst{(6)}}}
\def\7{{\sst{(7)}}}
\def\8{{\sst{(8)}}}
\def\9{{\sst{(9)}}}
\def\p{{\sst{(p)}}}
\def\q{{\sst{(q)}}}

\def\ssa{{\sst{(\alpha)}}}
\def\ssb{{\sst{(\beta)}}}
\def\ssg{{\sst{(\gamma)}}}
\def\j{{\sst{(j)}}}

\def\ve{\varepsilon}
\def\vf{\varphi}
\def\F{\Phi}
\def\wg{\wedge}

\def\thb{\bar{\theta}}
\def\Thb{\bar{\Theta}}
\def\barp{\bar{p}}
\def\barq{\bar{q}}
\def\barc{\bar{c}}
\def\bard{\bar{d}}
\def\e{\epsilon}

\def \bi{\bibitem}
\def \la {\label}

\def \l {\lambda}
\def\foot{\footnote}
\def \tl  {{\tilde \l}}
\def \sql {{\sqrt \l}}
\def \adss {$AdS_5 \times S^5$\ }
\newcommand{\rf}[1]{(\ref{#1})}
\def \ov {\over}

\def\Th{\Theta}
\def\vth{\vartheta}
\def\btheta{{\bar\theta}}
\def\ttheta{{{\tilde\theta}}}
\def\bttheta{{{\bar\ttheta}}}
\def\vth{\vartheta}

\def\ra{\rightarrow}
\def\N{{\cal N}}
\def\uM{\underline{M}}
\def\uA{\underline{A}}
\def\uN{\underline{N}}
\def\uP{\underline{P}}
\def\ua{\underline{a}}
\def\ub{\underline{b}}
\def\uc{\underline{c}}
\def\ud{\underline{d}}
\def\ue{\underline{e}}
\def\uf{\underline{f}}
\def\ui{\underline{i}}
\def\uj{\underline{j}}
\def\uk{\underline{k}}
\def\ual{\underline{\alpha}}
\def\ube{\underline{\beta}}
\def\um{\underline{m}}
\def\un{\underline{n}}
\def\up{\underline{p}}
\def\uq{\underline{q}}
\def\ur{\underline{r}}
\def\us{\underline{s}}

\def\umu{\underline{\mu}}
\def\unu{\underline{\nu}}
\def\ula{\underline{\l}}
\def\uka{\underline{\k}}
\def\usi{\underline{\s}}
\def\urh{\underline{\r}}
\def\cc{\circ}
\def\eqv{\equiv}

\def\ni{\noindent}

\def\Ep{E^{{}^{(+)}}}
\def\Em{E^{{}^{(-)}}}

\def\Mp{M^{{}^{(+)}}}
\def\Mm{M^{{}^{(-)}}}

\def \ha{{1\ov 2}}

\def\r{\rho}

\def\Y{{\rm Y}}
\def\X{{\rm X}}
\def\tY{\tilde{\rm Y}}
\def\tX{\tilde{\rm X}}
\def\dY{\dot{\rm Y}}
\def\dX{\dot{\rm X}}

\def \J {\mathcal{J}}
\def \del {\partial}

\def\dF{\dot{F}}
\def\dG{\dot{G}}
\def\df{\dot{f}}
\def\dx{\dot{x}}
\def \E {{\cal E}}
\def \S {{\cal S}}
\def \J {{\cal J}}

\def\ms{\mathcal{S}}
\def\mj{\mathcal{J}}
\def\soj{\fr{\ms}{\mj}}
\def \R {{\bf R}}
\def \om {\omega}
\def \bE {\bar E}
\def \x {{\cal X}}

\def \bi{\bibitem}
\def \la {\label}

\def \l {\lambda}
\def\foot{\footnote}
\def \tl  {{\tilde \l}}
\def \sql {{\sqrt \l}}
\def \adss {$AdS_5 \times S^5$\ }
\def \ov {\over}

\def \varpi {{\rm w}}

\def\thb{\bar{\theta}}
\def\Thb{\bar{\Theta}}
\def\mb{\bar{\m}}
\def\ab{\bar{\a}}
\def\zb{\bar{z}}
\def\psib{\bar{\psi}}
\def\barp{\bar{p}}
\def\barq{\bar{q}}
\def\barc{\bar{c}}
\def\bard{\bar{d}}
\def\e{\epsilon}
\def\wb{\bar{w}}
\def\lb{\bar{\l}}
\def\Jb{\bar{J}}
\def\Nb{\bar{N}}
\def\Zb{\bar{Z}}
\def\pab{\bar{\pa}}

\def\At{\tilde{A}}
\def\Bt{\tilde{B}}
\def\Ct{\tilde{C}}
\def\Dt{\tilde{D}}
\def\Et{\tilde{E}}
\def\Ft{\tilde{F}}
\def\Gt{\tilde{G}}
\def\Ht{\tilde{H}}
\def\Mt{\tilde{M}}
\def\Rt{\tilde{R}}
\def\at{\tilde{a}}
\def\bt{\tilde{b}}
\def\ct{\tilde{c}}
\def\dt{\tilde{d}}
\def\et{\tilde{e}}
\def\ft{\tilde{f}}
\def\gt{\tilde{g}}
\def\mt{\tilde{\mu}}
\def\nt{\tilde{\nu}}

\def\asth{\hat{*}}
\def\phh{\hat{\phi}}

\def\bA{{\bf A}}

\def\ola{\overleftarrow}
\def\ora{\overrightarrow}
\def\alt{\tilde{\a}}

\def\eh{\hat{e}}
\def\eph{\hat{\e}}
\def\ph{\hat{p}}
\def\alh{\hat{\a}}
\def\beh{\hat{\b}}
\def\gah{\hat{\g}}
\def\Fh{\hat{F}}
\def\muh{\hat{\m}}
\def\nuh{\hat{\n}}
\def\thh{\hat{\th}}
\def\dh{\hat{d}}
\def\ih{\hat{i}}
\def\jh{\hat{j}}
\def\kh{\hat{k}}
\def\deh{\hat{\d}}
\def\wh{\hat{w}}
\def\lah{\hat{\l}}
\def\Ah{\hat{A}}
\def\Ch{\hat{C}}
\def\Omh{\hat{\Omega}}

\def\xh{\hat{x}}

\def\ps{\rlap{\, /}\;\,p }
\def\ks{\rlap{\, /}\;\,k }

\def\gym{g_{YM}}

\def\adot{\dot{a}}
\def\bdot{\dot{b}}
\def\bpa{\bar{\pa}}

\def\pr{\prime}
\def\ssk{\medskip}
\def\bsk{\bigskip}

\def\clb{\color{blue}}
\def\clr{\color{red}}
\def\clv{\colo{violet}}

\def\t{\tau}
\def\cM{\mathcal{M}}
\def\S{\Sigma}

\def\N{\nabla}

\def\cR{\mathcal{R}}
\def\cL{\mathcal{L}}

\def\hb{\hbar}
\def\an{\hat{a}}
\def\ac{\hat{a}^\dag}
\def\hp{\hat{p}}

\def\Ec{{\cal E}}

\begin{document}


\title{
{\Large{\bf 
Scattering on Quasi-Spherical Black-Holes: Features and Beyond}}
}
{
\author{A.M. Arslanaliev$\,^{\vardiamondsuit,\spadesuit}$\footnote{arslanaliev.kh@gmail.com}
\,\, \,and  \,A.J. Nurmagambetov$\,^{\spadesuit,\vardiamondsuit,\varheartsuit}$\footnote{ajn@kipt.kharkov.ua}
\\ \\
$\,^{\vardiamondsuit}${ \it {\normalsize Department of Physics \& Technology, Karazin Kharkov National University,}}\\
{ \it {\normalsize 4 Svobody Sq., Kharkov 61022 UA} }
\\
$\,^{\spadesuit}${ \it {\normalsize Akhiezer Institute for Theoretical Physics of NSC KIPT}}\\
{ \it {\normalsize 1 Akademicheskaya St., Kharkov 61108 UA} }
\\
$\,^{\varheartsuit}${ \it {\normalsize Usikov Institute of Radiophysics and Electronics}}\\
{ \it {\normalsize 12 Ak. Proskury, Kharkov 61085 UA} }
}

\date{}

\maketitle

\abstract{
Recent developments in the gravitational waves interferometry require more pertinent theoretical models of gravitational waves generation and propagation. Untouched possible mechanisms of spin-2 spacetime perturbations production, we will consider their subsequent scattering on other black holes (BHs). Specifically, we consider a generalization of the Regge-Wheeler-Zerilli equations for the case of distorted BHs (BHs surrounded with matter) in Minkowski and Anti-de Sitter spacetimes, the~metric potential of which obeys the Liouville equation. We establish significant differences in scattering characteristics of waves of different spins and angular momenta, including the gravitational waves, caused by losing the spherical symmetry of their propagation background. In~particular, we demonstrate the strong impact of the background geometry deformation on the grey-body factors, hence on the absorption cross-sections of scattering waves, and~explore the issue of stability of the background geometry upon changing the deformation degree~parameters. 
}

\newpage

\section{Introduction}

Progress in observation astronomy is going so fast that we become beholders of transforming astrophysics into genuinely physical discipline, where theoretical predictions are supervised with experimentally verified data. A~lot of discoveries in the field were done since the century beginning. We will outline just a few breakthroughs in the contemporary multi-messenger Astrophysics~\cite{Bartos:2017}, related in different ways to black holes (BHs):
\begin{itemize}
\item
First, we have the {LIGO}
-Virgo scientific collaborations to mention for their fundamental contribution in registrations of gravitational waves (GWs)~\cite{TheLIGOScientific:2016pea}. It is believed that the GWs are mostly induced in processes which involve BHs.
\item
Second, it is important to notice the detection of ultra high energy cosmic rays (UHECRs) of energy $\sim$$10^{19}$ eV---that far exceeds energies of the LHC
   (Large Hadron Collider)---coming from active galactic nuclei (AGN) \cite{Abraham:2007bb}. It is believed that super-massive BHs (SMBHs) are located in the core of the AGN~\cite{Eckart:1996zz,Ghez:1998ph}, and~mechanisms of the UHECRs generation are directly related to BH physics; see, e.g., refs.~\cite{Felice:2003,Felice:2004,Kachelriess:2008ze,Stanev:2010ys,Dawson:2017rsp,Park:2017dib,Nurmagambetov:2018het,Guepin:2017abw,Nurmagambetov:2019mih,Nurmagambetov:2019bqz,Comisso:2019frj,Nurmagambetov:2020dbc,Nurmagambetov:2020ann,Comisso:2020ykg}, in this respect.
\item
\emergencystretch 2em

Next, it is worth mentioning achievements of the Event Horizon Telescope (EHT) collaboration in revealing the BH event horizon (for M87 SMBH with the mass \mbox{6.5 $\times$ $10^9$~$M_\odot $ \cite{Akiyama:2019cqa}}). Due to the EHT activity it is believed more and more that Astrophysical BHs (BHs in the sky) share common properties with mathematical BHs (solutions to Einstein equations).
\item
Last but not least: the recent data analysis of the EHT team puts the Einstein theory on the test~\cite{Psaltis:2020prl}. It turns out that general relativity works well, and~many alternative theories of gravity should be discarded as that of contradicting experimental data. However, one has to notice that the conclusions of ref.~\cite{Psaltis:2020prl} cannot be considered as ultimate ones; see, e.g.,~ref.~\cite{Gralla:2020pra}, in this respect.
\end{itemize}

The mentioned achievements in contemporary astrophysics show how various studies of compact astrophysical objects give new prospects in resolving old and revealing new problems, which impact the development of Physics at~whole.

A large enough amount of information on BHs comes with d-waves (spin 2 waves; GWs). In~view of this fact, any current studies of d-waves become potentially important. One of the branches in this activity is scattering d-waves on compact astrophysical objects. Starting from first seminal papers on scattering of different spin particles on Kerr and Schwarzschild BHs~\cite{Starobinskil:1974nkd,Fabbri:1975sa,Unruh:1976fm,Sanchez:1976fcl,Sanchez:1976xm,Sanchez:1977si,Sanchez:1977vz}, it has been obtained, analytically and numerically, cross-sections and grey-body factors of external particle fluxes propagating in effective potentials of various BH models; see, e.g., refs.~\cite{Cvetic:1997xv,Klebanov:1997cx,Cvetic:1997ap,Kanti:2004nr,Grain:2005my,Harmark:2007jy,Boonserm:2008zg,Li:2009zzf,Keshet:2007be,Gonzalez:2010vv,Boonserm:2013dua,Dong:2015qpa,Catalan:2014ama,Gray:2015xig}.

In what follows, we will be interested in solving to the scattering problem for the GWs in the background of the so-called distorted BHs. Rationales behind appearing the distorted BHs on the scene come from the point that the Schwarzschild solution is highly idealized and cannot be fully employed in any real astrophysical problem; the same concerns the Kerr solution. Any presence of matter does not just change the metric, but~may drastically change its type, due to the backreaction of matter fields enforcing the spacetime geometry to evolve in time. However, time-dependent solutions to the Einstein equations in analytical form are extremely rare (and for very specific matter fields) \cite{Stephani:2003,Griffiths:2009} and~generally require involved numerical simulations. A~significant simplification occurs upon replacing the backreaction of matter with an effective distortion of a static/stationary BH metric with keeping the time-independence of the solution. Then, the~solution describing a static distribution of matter localized outside the BH horizon and the vacuum spacetime at its proximity is called the distorted BH~\cite{GH:1982,Weyl:1917}; also see refs.~\cite{Thorne:1970,Thorne:1972ji}. 
There are various astrophysical applications for the distorted BH-type metrics (see, e.g.,~refs. \mbox{\cite{Tomimatsu:1984sx,Breton:1997,Breton:1998sr,Semerak:2000,Zacek:2002,Letelier:2003ea,Shoom:2015slu,Kunz:2017mfe}}), among~which one can find the effective description of a double (neutron) ``star'' system, when the tidal forces of one of the components deforming the shape of the other are mimicked in the solution~\cite{Araujo:1995vb,Araujo:1997sm,Semerak:2016gfz}.

In the next parts of the paper, we will explore the scattering of d-waves on another class of distorted BHs, in which spacetime configuration is defined by a specific solution originally proposed and studied in refs.~\cite{Moskalets:2014hoa,Moskalets:2015xxa,Moskalets:2017koi,Moskalets:2016uno} and later rediscovered in ref.~\cite{Boos:2017pyd} in a different context. In~Section~\ref{sec:2}, we briefly discuss the relationship between the standard metric of axisymmetric distorted BH~\cite{GH:1982,Weyl:1917} and the quasi-spherical BH solution of refs.~\cite{Moskalets:2014hoa,Moskalets:2015xxa,Moskalets:2017koi,Moskalets:2016uno}. In~its customary formulation, the~Weyl solution includes two metric potentials (functions of the radial direction and polar angle) and describes a family of distorted BHs in dependence on their specific choice. The~proposed generalization of the Weyl solution with third metric potential, in Section~\ref{sec:2}, now depending on two angles, breaks the original axial symmetry. It results in fixing two metric potentials of the Weyl solution, so that the final geometry becomes that of a quasi-spherical BH, the~metric potential of which obeys the Liouville equation. The~quasi-spherical solution also describes a family of BHs, now determined by the choice of a single metric potential.
The local spacetime geometry we will explore further on is that of a static neutral distorted BH. It is evidently not enough to describe real astrophysical problems in full extent (see more on a quasi-spherical generalization of the Kerr solution in the last section). However, it is sufficient to reveal main features in the scattering processes of external particle fluxes, which will also be inherent in scattering by rotating distorted black~holes.

In Section~\ref{sec:3}, we consider equations for gravitational perturbations over backgrounds of quasi-spherical neutral BHs in Minkowski and anti-de Sitter (AdS) spacetimes and examine the issue of separation of variables. As~it is well known, a~possibility to separate the variables in dynamical equations of relativistic fields in a curved background is related to symmetry properties of the background spacetime. By~use of the Newman-Penrose formalism, we figure out the spacetime of quasi-spherical BHs is of type D in the Petrov classification. Therefore, the~radial and angular parts of equations for small perturbations over the quasi-spherical BH background can be set apart, and~the resulting expressions generalize the Regge-Wheeler-Zerilli equations and the spherical harmonics~equation. 

In Section~\ref{sec:4}, we focus on the angular part of the scattering problem and survey differences in solution to the angular differential equation for the distorted/quasi-spherical spacetime in compare to the standard Schwarzschild background. In~general, the~solution to the angular equation in a quasi-spherical BH background is reduced to the spectral problem for infinite-dimensional matrices. For~the background with axial symmetry, viable in many astrophysical problems, we can make further simplifications, and~to solve the spectral problem numerically. Here, we find the principle difference between the quasi-spherical and spherically-symmetric cases, which will have the foremost impact on the scattering process: the eigenvalues of the examined spectral problem are not integers anymore; for each scattering mode, there is a set of eigenvalues, the number of which is determined by the corresponding to spherical symmetry value of the scattering mode angular momentum and by its projections (i.e., by the set of $(l,m)$); and finally, the~eigenvalues depend on the deformation degree of the background geometry from spherically-symmetric, specified by a single parameter. On~account of numerical computations we recovered the functional dependence of the generalized eigenvalues on the degree of deformation in the axially-symmetric~case.

Section~\ref{sec:5} contains computations of the grey-body factor for different scattering waves on a quasi-spherical Schwarzschild BH and further comparison of the regular spherically-symmetric case~\cite{Gray:2015xig} to its counter-part with a non-trivial deformation. Because~the generalized eigenvalues found in the preceding section carry on two indices and enter the Regge-Wheeler-Zerilli equations via the separation constant, numerically computed grey-body factors for each type of perturbations---scalar, vector, and tensor---also become differentiated by $(l,m)$ indices. Explicitly, for~every scattering mode with the angular momentum $l$, we find $l+1$ different values of the grey-body factor, properties of which were compared to that of the spherically-symmetric case. In~particular, we find that the grey-body factors, as~functions of the deformation degree, increase with increasing the value of this parameter, as well as~that the transparency of the effective potential is reached for $(l,l)$ scattering modes at the lowest value of corresponding~frequencies.

In Section~\ref{sec:6}, we consider the issue of stability of BH backgrounds by studying the quasinormal modes (QNMs). Here, we briefly review the relation between the stability of a spacetime and positivity of effective potentials in the Regge-Wheeler-Zerilli equations. Since the effective potential of small perturbations over distorted/quasi-spherical BH backgrounds depends on the deformation degree, this parameter becomes crucial for determining the stability of the background geometry against small perturbations. We find that the value of the deformation degree equal to one is the critical value for the stability of an axially-symmetric quasi-spherical Schwarzschild BH in Minkowski and AdS spacetimes. And,~if the result does not depend on the size of a BH in flat spacetime, the~instability of a Schwarzschild-AdS BH may only be encountered for the so-called large BHs, the~event horizons of which are of the next order in compare to the characteristic scale of empty AdS~space.

Finally, discussion of the results and summary of our findings are collected in the last~section.


\section{Background Metric: From Distorted to Quasi-Spherical Black~Hole}\label{sec:2}


Let us begin with a clarification of how the background metric, mainly used throughout the paper, is related to the metric of a distorted~BH.

A distorted BH solution to the flat spacetime Einstein vacuum equations is traditionally described by the Weyl axisymmetric metric \cite{Weyl:1917} in the cylindrical space-time coordinates $(t,\r,\th,\vf)$ \cite{GH:1982}, $t$ is the time coordinate. However, when the case is about a static axisymmetric solution with an arbitrary quadrupole moment, things get essentially simplified in the prolate spheroidal space-time coordinates $(t,x,y,\vf)$ \cite{Erez:1959}, in~which the line element looks as follows: 
\be
ds^2=-e^{2\psi(x,y)}dt^2+M^2 e^{-2\psi(x,y)}\Bigg[e^{2\g(x,y)}(x^2-y^2)\left(\fr{dx^2}{x^2-1}+\fr{dy^2}{1-y^2} \right)+(x^2-1)(1-y^2)d\vf^2 \Bigg].
\la{ERds2}
\ee

\noindent To obey the {Einstein equations} in vacuum, the metric potentials $\y(x,y)$ and $\g(x,y)$ of \rf{ERds2} should fulfill the following set of equations~\cite{Erez:1959,Quevedo:1989}: 
\be
\pa_x\left((x^2-1)\pa_x \y\right)+\pa_y\left((1-y^2)\pa_y \y \right)=0,
\la{psieqxy}
\ee
\be
\pa_x \g=\fr{1-y^2}{x^2-y^2}\left[x(x^2-1)(\pa_x \y)^2-x(1-y^2)(\pa_y\y)^2-2y(x^2-1)\pa_x\y \pa_y\y \right],
\la{paxgxy}
\ee
\be
\pa_y \g=\fr{x^2-1}{x^2-y^2}\left[y(x^2-1)(\pa_x \y)^2-y(1-y^2)(\pa_y \y)^2+2x(1-y^2)\pa_x\y\pa_y\y \right],
\la{paygxy}
\ee
where $\partial_a \equiv \pa/\pa a$.

{Equations} \rf{psieqxy}--\rf{paygxy} leave enough freedom in choosing the specific form of the metric potentials. The~entering \rf{ERds2} constant $M$ is associated to the mass of the~BH. 

A non-trivial generalization of \rf{ERds2}, which includes the third metric potential $\Phi$, now dependent on 
$(y,\vf)$ coordinates,~therefore breaking the axisymmetric invariance of the original metric, is
\[
ds^2=-e^{2\psi(x,y)}dt^2+M^2 e^{-2\psi(x,y)} \times
\]
\be
\times\Bigg[e^{2\g(x,y)}(x^2-y^2)\left(\fr{dx^2}{x^2-1}+\fr{e^{\Phi(y,\vf)}dy^2}{1-y^2} \right)+e^{\Phi(y,\vf)}(x^2-1)(1-y^2)d\vf^2 \Bigg].
\la{ERds2gen}
\ee

In this case, the presence of the third metric potential with the specific dependence on coordinates puts strong restrictions on the metric potentials $\y(x,y)$ and $\g(x,y)$. In~particular, the~non-triviality of $\Phi(y,\vf)$ results in the appearance of non-diagonal terms in the Ricci tensor that, in~the absence of matter in the Einstein equations, sets the following constraints on $\g(x,y)$:
\be
\pa_x\g=\fr{x(1-y^2)}{(x^2-y^2)(x^2-1)},\qquad \pa_y \g=\fr{y}{x^2-y^2},
\la{geqsL}
\ee
the further account of which leads to the additional restriction on $\y(x,y)$:
\be
\pa_x \y\, \pa_y \y=0.
\la{psieqL}
\ee 

A general solution to \rf{geqsL} comes as follows:
\be
\g(x,y)=\fr12\ln \fr{x^2-1}{x^2-y^2}+C_\g;
\la{gsolL}
\ee
then, solving for the corresponding equation for $\y(x,y)$ constrained by \rf{psieqL}, we get
\be
\y=\fr12\ln\fr{x-1}{x+1}+C_\y.
\la{psisolL}
\ee

Note that \rf{psisolL} is a particular solution for $\y$, which is convenient to choose in this form to establish in what follows the relation of \rf{ERds2gen} to a quasi-spherical Schwarzschild BH metric. With~$\g$ and $\y$ of \rf{gsolL}, \rf{psisolL}, the~metric potential $\Phi(y,\vf)$ is confined by the following differential~equation:
\be
\pa_y\left((y^2-1)\pa_y\Phi \right)-2(e^\Phi-1)+e^{2 C_\g}\fr{\pa^2_\vf \Phi}{y^2-1}=0.
\la{PhieqL}
\ee

Fixing two integration constants $C_\g$ and $C_\y$ to be zero and introducing new coordinates $(r,\th)$, related to $(x,y)$ via
\be
x=\fr{r}{M}-1,\qquad y=\cos\th,
\la{rthdef}
\ee
we recover the metric of a neutral BH~\cite{Moskalets:2016uno,Moskalets:2017koi} (also see ref.~\cite{Boos:2017pyd}) in the spherical coordinates
\be
ds^2=-f(r)dt^2+\fr{dr^2}{f(r)}+r^2 e^{\chi({\th}, \vp)}\left( d\th^2+\sin^2\th\, d\vp^2 \right),
\label{metric}
\ee
with the standard red-shift factor $f(r)=1-2M/r$ and the ``smearing'' function of the BH horizon $\chi(\th,\vf)$, which follows from $\Phi(y,\vf)$ after the coordinate transformations \rf{rthdef}. Equation \rf{PhieqL} turns into the spherical Liouville equation
\be
\D_{\th,\vp}\,\chi(\th,\vp) + 2(e^{\chi(\th,\vp)}-1) = 0,
\la{Lspheq}
\ee
with
\be
\D_{\th,\vp} = \frac{1}{\sin\theta}\frac{\partial}{\partial\theta}\sin\theta\frac{\partial}{\partial\theta} + \frac{1}{\sin^2\theta}\frac{\partial^2}{\partial\varphi^2} .
\label{Laplac}
\ee 

Recall that the~Liouville equations are exactly-solvable and possess general analytic solutions in terms of unrestricted functions; see, e.g., refs.~\cite{Crowdy:1997,Popov:1993}, or~a brief summary in refs.~\cite{Moskalets:2014hoa,Moskalets:2016uno} and Appendix A in below. Therefore, we have enough freedom in choosing a function to fit the desired shape of a two-dimensional surface. For~a BH, this surface is a quasi-spherical/distorted horizon; in the case of a rigid celestial body (as a neutron star), the two-dimensional surface is a 2-dimensional (2D) slice of the 3-dimensional (3D) surface, deformed by the tidal forces of another star/BH.

Therefore, we have established the generalization of the standard Weyl-Erez-Rosen distorted BH solution \rf{ERds2} with additional, breaking the axial symmetry, metric potential (cf. \rf{ERds2gen}). As~a result of such modification, the~solution becomes more rigid, and~corresponds, after~turning to the standard spherical coordinates, to~the neutral BH solution with the Liouville mode~\cite{Moskalets:2014hoa,Moskalets:2015xxa,Moskalets:2017koi,Moskalets:2016uno}. That makes possible to consider the BH solution of refs.~\cite{Moskalets:2014hoa,Moskalets:2015xxa,Moskalets:2017koi,Moskalets:2016uno,Boos:2017pyd} {as a specific distorted BH}. Taking this point of view, we will not differentiate further on distorted and quasi-spherical BHs and will freely use both terms on equal~footing.

To sum up this part of the work, the~spacetime background we will consider throughout the paper is defined by the line element
\begin{equation}
ds^2 = -f(r)dt^2  + \frac{dr^2}{f(r)} + r^2e^{\chi(\theta,\varphi)}(d\theta^2+\sin^2\theta d\varphi^2),
\label{ds2def}
\end{equation}
with 
\be
f(r) = \frac{\Delta}{r^2}, \quad  \Delta = r^2 - 2Mr +\k^2 r^4 .
\la{fdef}
\ee

A non-trivial value {of} $\kappa^2=- {\Lambda}/{3}$, where $\kappa$ is the inverse characteristic length of AdS spacetime and $\Lambda$ is its cosmological constant, corresponds to the AdS-Schwarzschild BH; $\kappa=0$ is that of the flat spacetime. To~solve the vacuum Einstein equations, the metric potential $\chi(\theta,\varphi)$ has to be confined to the spherical Liouville Equation \rf{Lspheq} with the angular Laplacian \rf{Laplac}.

\section{Basic Equations and Separation of~Variables}\label{sec:3}

Technically, we would like to extend the known Regge-Wheeler-Zerilli~\cite{Regge:1957td,Zerilli:1970se} equations for small d-wave perturbations over the background (\ref{ds2def}) and to solve them. First step on this way is to separate the variables in the corresponding relativistic spin equation. Our previous experience with s-wave perturbations over such a background~\cite{Moskalets:2016uno} indicates the possibility to separate variables in the massless Klein-Gordon equation, the~explicit form of which in the background (\ref{ds2def}) is as follows:
\begin{equation}
-\frac1{f}\partial^2_t\Phi+\frac1{r^2}\partial_r(r^2 f\,\partial_r \Phi)+\frac{e^{-\chi(\theta,\varphi)}}{r^2}\triangle_{\theta,\varphi}\,\Phi=0,
\label{KG}
\end{equation}
by use of the separation ansatz
\begin{equation}
\Phi(t,r,\theta,\varphi)=e^{-i\omega t}\, Q_0(r)\,\Theta(\theta,\varphi).
\label{swavesep}
\end{equation}

{But} it is well-known that the (im)possibility to separate the variables strongly depends on special properties of the spacetime upon the consideration, specifically on~its type within the Petrov spacetime classification scheme~\cite{Petrov:1954,Petrov:1969book}.  

To determine the Petrov type of the metric in hand, we follow the Newman-Penrose (NP) formalism~\cite{Newman:1961qr} and introduce the null {tetrad}
\begin{equation}
e_{(1)}^\mu = l^\mu, \;\; e_{(2)}^\mu = n^\mu, \;\; e_{(3)}^\mu = m^\mu,\;\; e_{(4)}^\mu = \bar{m}^{\mu},
\label{NP}
\end{equation}
which is standardly related to the metric via $g_{\mu\nu}=e_\mu^{(a)}\eta_{(a)(b)} e_\nu^{(b)}$ with 
\be
\eta_{(a)(b)}=\left(
\begin{array}{cccc}
0&-1 &0&0\\
-1 &0&0&0\\
0&0&0&1\\
0&0&1&0
\end{array}
\right).
\la{etadef}
\ee
The curved space indices $\m,\n$ run over $0,1,2,3$; ''$0$'' corresponds to time direction. The latin indices ''$(a)$'' are that of flat tangent space and run over $1,2,3,4$.

Explicitly, for~the metric (\ref{ds2def}), we {get} 
\begin{equation}
l_{\mu} = \delta_{\mu 0} - \frac{\delta_{\mu r}}{f(r)}, \;\; n_\mu = \frac{f(r)}{2}\,\delta_{\mu 0} + \frac{\delta_{\mu r}}{2}, \;\; m_\mu = \frac{r}{\sqrt{2}}\,e^{\frac{\chi(\theta,\varphi)}{2}}(\delta_{\mu\theta} + i\sin\theta\delta_{\mu\varphi}),
\label{NPdef}
\end{equation} 
where $\d_{\m\n}$ is the Kronecker delta.

By use of the NP tetrad (\ref{NPdef}), one may compute various coefficients of the spin-connection $\gamma_{(c)(a)(b)} = e_{(c)}^\nu e_{(a)\nu;\mu}e^\mu_{(b)}$, where, as usual, the semicolon corresponds to the covariant derivative over the metric $g_{\m\n}$, {and} observe vanishing the coefficients (in the notation of ref.~\cite{Newman:1961qr}) $\kappa$, $\sigma$, $\nu$, $\lambda$. According to the Goldberg-Sachs theorem, trivialization of $\kappa$, $\sigma$, $\nu$, $\lambda$ corresponds to the Petrov type D metrics~\cite{Petrov:1954,Petrov:1969book,Newman:1961qr}. 

Since the pioneering papers by Teukolsky~\cite{Teukolsky:1973ha,Press:1973zz}, it was realized that, in the Petrov type D spacetimes, equations for gravitational perturbations decouple for quantities
\begin{equation}
\psi_0 = -C_{\alpha\beta\gamma\delta}l^{\alpha}m^\beta l^\gamma m^\delta, \qquad \psi_4 = -C_{\alpha\beta\gamma\delta}n^{\alpha}\bar{m}^\beta n^\gamma \bar{m}^\delta ,
\label{psi04def}
\end{equation}
forming with the Weyl tensor $C_{\alpha\beta\gamma\delta}$ and the NP tetrad (\ref{NPdef}). As~we will see, in short, these quantities correspond to the odd (axial) gravitational perturbation of ref.~\cite{Regge:1957td} and even (polar) d-wave perturbation of ref.~\cite{Zerilli:1970se}. 

Indeed, by~use of the {ansatz} (we use indices ''$+2$'' and ''$-2$'' for odd and even d-wave perturbations, respectively)
 \begin{equation}
\left(
\begin{array}{c}
\psi_0\\ 
r^4 \psi_4
\end{array}
\right)= e^{-i \omega t} \left(
\begin{array}{c}
\Psi_{+2}(r) \Theta_{+2}(\theta,\varphi)\\ 
\Psi_{-2}(r) \Theta_{-2}(\theta,\varphi)
\end{array}
\right),
\label{psi04sep}
\end{equation}
one may separate the temporal, radial and angular parts of the gravitational perturbations over the basic metric (\ref{ds2def}). Upon~the separation of the angular part of the d-wave perturbations, we arrive at the following master equation for the fundamental angular variable $\Theta(\theta,\varphi)$:
\begin{equation}
\Delta_{\theta,\varphi}\Theta(\theta,\varphi) + Ce^{\chi(\theta,\varphi)}\Theta(\theta,\varphi) = 0,
\label{Thetaeq}
\end{equation}
in which, for~the sake of convenience in further comparing to the spherically symmetric example, we set the separation constant to $C=\nu(\nu+1)$. Apparently, the~spherical symmetry of the background spacetime is recovered for the trivial metric potential $\chi(\theta,\varphi)=0$ ($\chi(\theta,\varphi)=\mathrm{const}$ also gets, re-scaling the radial coordinate, the~spherical symmetry back). However, in~general, the~spherical symmetry is lost, that means $\nu$ does not fall into the set of integers. The~fundamental angular variable $\Theta(\theta,\varphi)$, together with angular differential operators
\begin{equation}
L_n = \partial_\theta - \frac{i}{\sin\theta}\partial_\varphi + n\left(\cot\theta + \frac12\left(\partial_\theta\chi - \frac{i}{\sin\theta}\partial_\varphi\chi\right)\right),
\label{Lndef}
\end{equation}
{form} $\Theta_{\pm 2}(\theta,\varphi)$ of (\ref{psi04sep}): 
\[
\Theta_{+2}(\theta,\varphi) = e^{-\frac12\chi(\theta,\varphi)}L^\dagger_{-1}e^{-\frac12\chi(\theta,\varphi)}L^\dagger_0\,\Theta(\theta,\varphi), 
\]
\begin{equation}
\Theta_{-2}(\theta,\varphi) = e^{-\frac12\chi(\theta,\varphi)}L_{-1}e^{-\frac12\chi(\theta,\varphi)}L_0\,\Theta(\theta,\varphi).
\label{Thetapm2}
\end{equation}

For the radial part of the perturbation equations of $\psi_0$ and $\psi_4$, the NP formalism is equivalent (after transition to new radial functions $Q_{\pm 2}(r)$; see refs.~\cite{Chandr,Otsuki91} for details) to the Regge-Wheeler-Zerilli equations
\begin{equation}
\left[\frac{\partial^2}{\partial r_*^2}  + \omega^2 - V_s(r)\right] Q_{s} = 0,  \;\;r_* \in(-\infty, +\infty), \;\; s=\pm 2 .
\label{RWZeqs}
\end{equation}

Here, $r_*$ is the so-called ``tortoise'' coordinate ${dr}/{dr_*} = f(r)$; $V_{\pm 2}$ are the effective potentials, entering either the Regge-Wheeler~\cite{Regge:1957td}
\begin{equation}
V_{+2}(r) = -\frac{3f(r)\partial_rf(r)}{r}  + \nu(\nu+1)\frac{f(r)}{r^2} + 6\kappa^2f(r),
\label{V+2}
\end{equation}
or the Zerilli~\cite{Zerilli:1970se}

\begin{equation}
V_{-2}(r)= \frac{2f(r)}{r^3}\frac{9M^3 + 3c^2Mr^2 + c^2(1+c)r^3 + 9M^2\left(cr+3  \kappa^2r^3\right)}{(3M + cr)^2},\quad c = \frac{\nu(\nu+1)}{2}-1
\label{V-2}
\end{equation}
\noindent equations. Altogether, the Regge-Wheeler-Zerilli Equations~(\ref{RWZeqs}) look like a stationary Schr\"odinger equation for the wave functions $Q_{\pm 2}$ and the effective potentials (\ref{V+2}) and \mbox{(\ref{V-2})}. Note that, due to the lack of the spherical symmetry, we do not have the total angular momentum (and its projection, as well) conservation, though~we can expect a quantization of the generalized angular momentum quantum numbers $\nu$.

Let us also emphasize that the scalar and vector perturbations over the basic metric \mbox{\rf{ds2def}} are described by a Schr\"odinger-like Equation \rf{RWZeqs}, as well, with~the effective potential \rf{Veffax}, in which one has to choose $s=0$ and $s=1$ for the scalar and vector modes, respectively. The~common separation ansatz for the scalar, vector, and tensor perturbations looks as follows (cf. \rf{psi04sep}):
\be
\left(
\begin{array}{c}
r^{-1}\Phi\\
r \tilde{\phi}\\
\y_0\\
r^4 \y_4
\end{array}
\right)=e^{-i\w t}\left(
\begin{array}{c}
Q_0(r)\, \Theta(\th,\vf)\\
Q_1(r)\, \Th_1(\th,\vf)\\
\Psi_{+2}\, \Th_{+2}(\th,\vf)\\
\Psi_{-2}\, \Th_{-2}(\th,\vf)
\end{array}
\right).
\la{psiallmodessep}
\ee

The $\Th(\th,\vf)$ function of \rf{psiallmodessep} obeys the master Equation \rf{Thetaeq}; $\Th_{\pm 2}(\th,\vf)$ are those of \rf{Thetapm2}. $\Th_{1}(\th,\vf)$ is also related to the fundamental angular variable $\Th(\th,\vf)$; in this case we have
\be
\Theta_{1} = e^{-\frac12\chi(\theta,\varphi)}L_1^\dagger\,\Theta(\theta,\varphi),
\la{Th1def}
\ee
with the angular differential operator $L_1$ of \rf{Lndef}. Therefore, the~``axial'' perturbations over the quasi-spherical BH background are described by equations
\begin{equation}
\left[\frac{\partial^2}{\partial r_*^2}  + \omega^2 - V_s(r)\right] Q_{s} = 0,  \;\;r_* \in(-\infty, +\infty), \;\; s=0,1,2,
\label{s012axeqs}
\end{equation}
with the effective potentials~\cite{Regge:1957td,Fabbri:1975sa,Unruh:1976fm,Sanchez:1976fcl,Horowitz:1999jd,Cardoso:2001bb,Cardoso:2003cj}
\begin{equation}
V_s(r) = \frac{(1-s^2)f\partial_rf}{r}  + \nu(\nu+1)\frac{f}{r^2} + 3s(s-1)\kappa^2f,\quad s=0,1,2.
\label{Veffax}
\end{equation}

The polar d-wave perturbation is still described by the Zerilli Equation \rf{RWZeqs} with $s=-2$ and with the effective potential \rf{V-2}. The~shape of the effective potentials for flat and AdS spacetimes in the spherically symmetric case are depicted in Figures~\ref{fig1} and~\ref{fig2}.

Two comments on Figures~\ref{fig1} and \ref{fig2} are in order. First, the~shape of the effective potentials in Minkowski space-time (both panels of Figure~\ref{fig1}) keeps its form up to spatial (radial in the case) infinity. It is not true for AdS spacetime, where the shape of the potentials (in both panels of Figure~\ref{fig2}) will change to $\sim$$r^2$ profile at $r \gg 1$. It, in particular, means that there are bound states in AdS spacetime that result in the discrete part of the spectrum of admissible perturbations~\cite{Moskalets:2016uno,Cardoso:2003cj,Burgess:1984ti,Harmark:2007jy,Konoplya:2011qq}. And~second, the~effective potentials of the Regge-Wheeler-Zerilli Equations \rf{V+2} and \rf{V-2} are almost the same as in Minkowski, as well as in AdS spacetimes (cf. right panels in Figures~\ref{fig1} and \ref{fig2}). This is the indication of isospectrality of the Hamiltonians, entering Equations \rf{RWZeqs} \cite{Cardoso:2003cj,Dias:2018etb,Glampedakis:2017rar,Moulin:2019bfh}.
\begin{figure}[ht]
\begin{minipage}[h]{0.49\linewidth}
\center{\includegraphics[width=1.\linewidth]{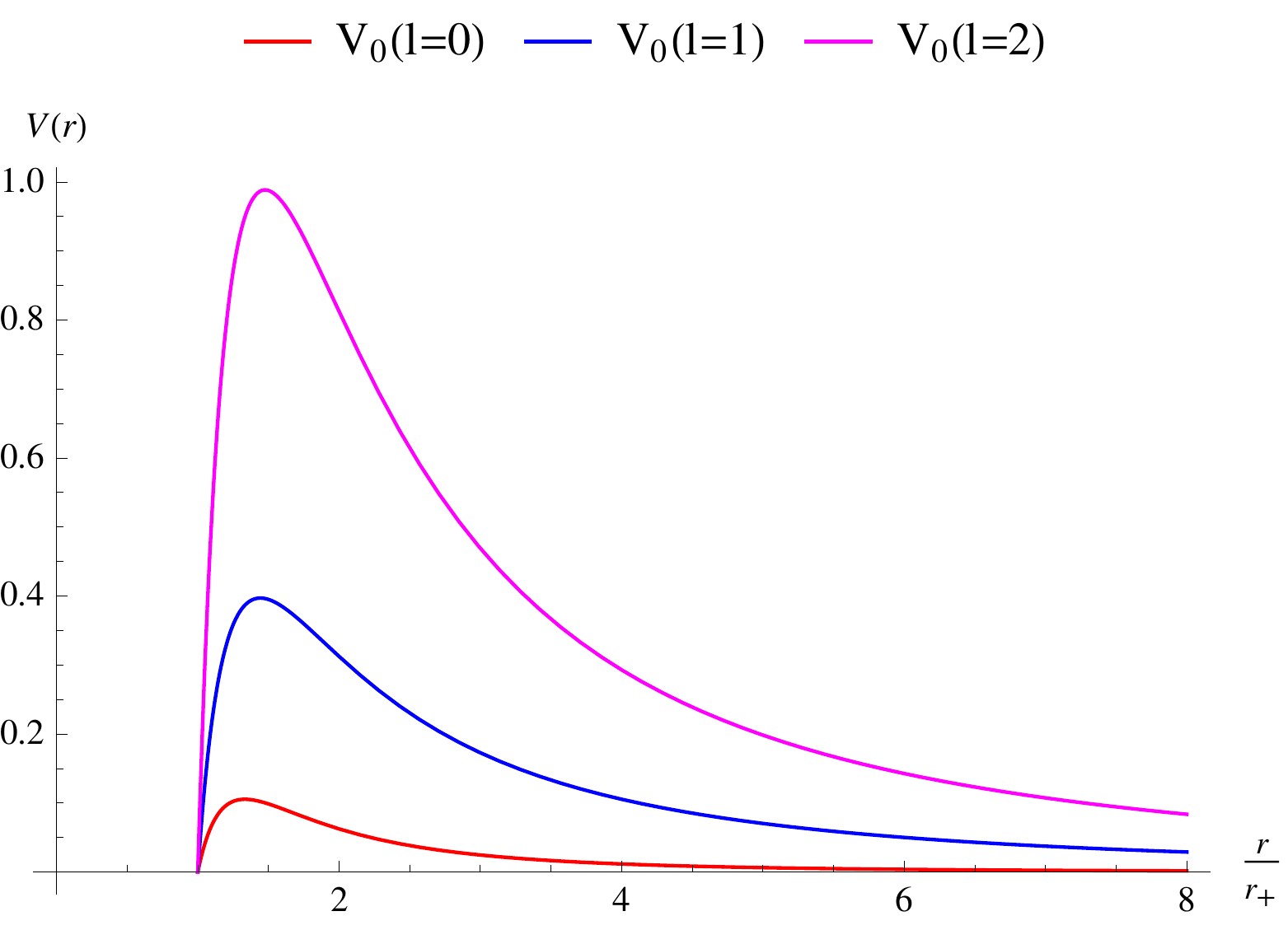} }
\end{minipage}
\hspace{0.29cm}
\begin{minipage}[ht]{0.49\linewidth}
\center{\includegraphics[width=1.\linewidth]{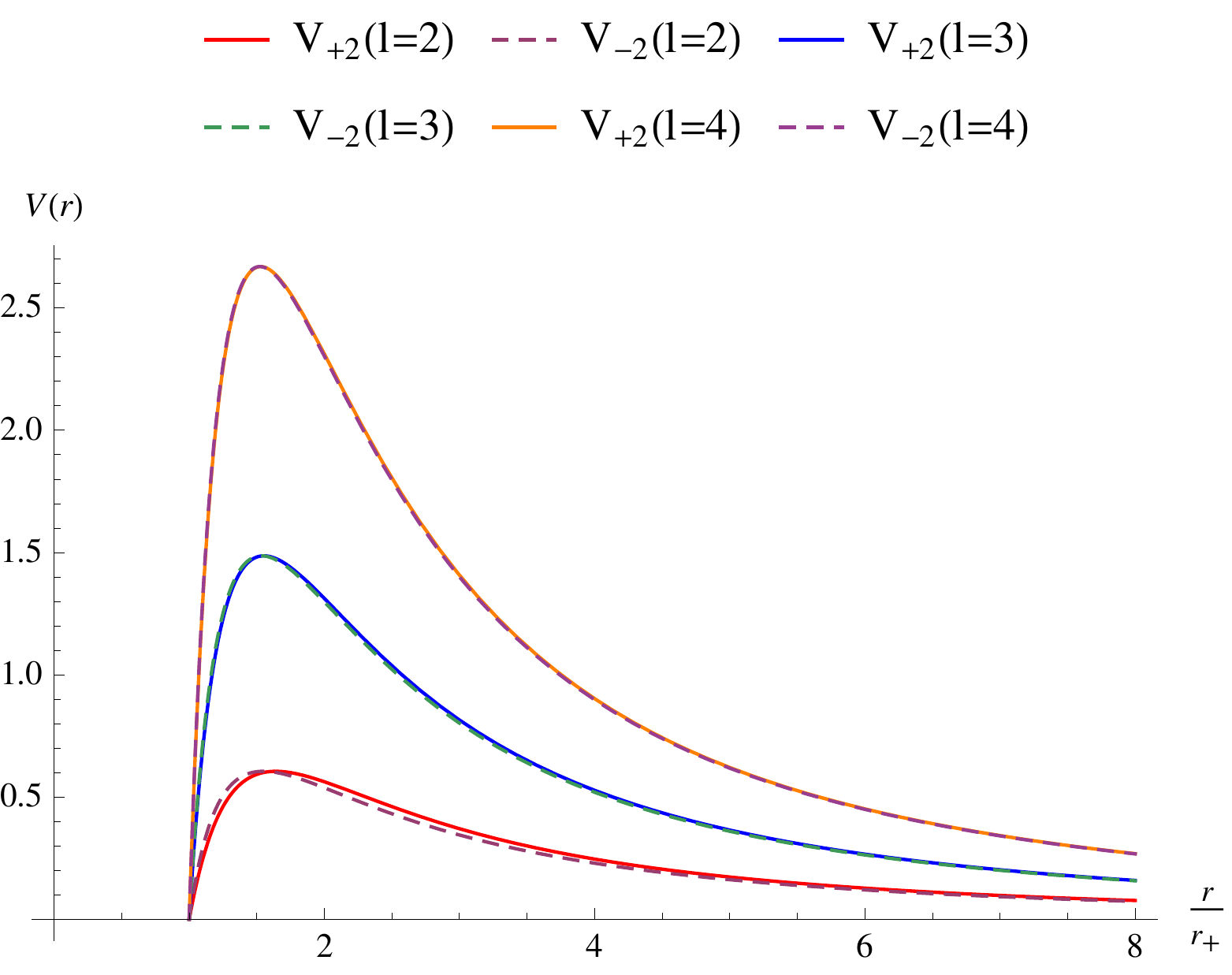}}
\end{minipage}
\caption{\label{fig1} \textbf{Left panel}: the shape of the effective potential of different modes ($l=0,1,2$) of the scalar perturbation over the spherically-symmetric Schwarzschild background. \textbf{Right panel}: comparison of the effective potentials $V_{+2}(r)$ (solid lines) and $V_{-2}(r)$ (dashed lines) of the d-wave perturbations ($l=2,3,4$) over the spherically-symmetric Schwarzschild background. The~(almost) coincidence of $V_{\pm 2}$ reflects the isospectrality of the corresponding Hamiltonians of \rf{RWZeqs}. The~value of the horizon radial location is chosen to be $r_+=1$.}
\end{figure} 

\begin{figure}[ht]
\begin{minipage}[h]{0.49\linewidth}
\center{\includegraphics[width=1.\linewidth]{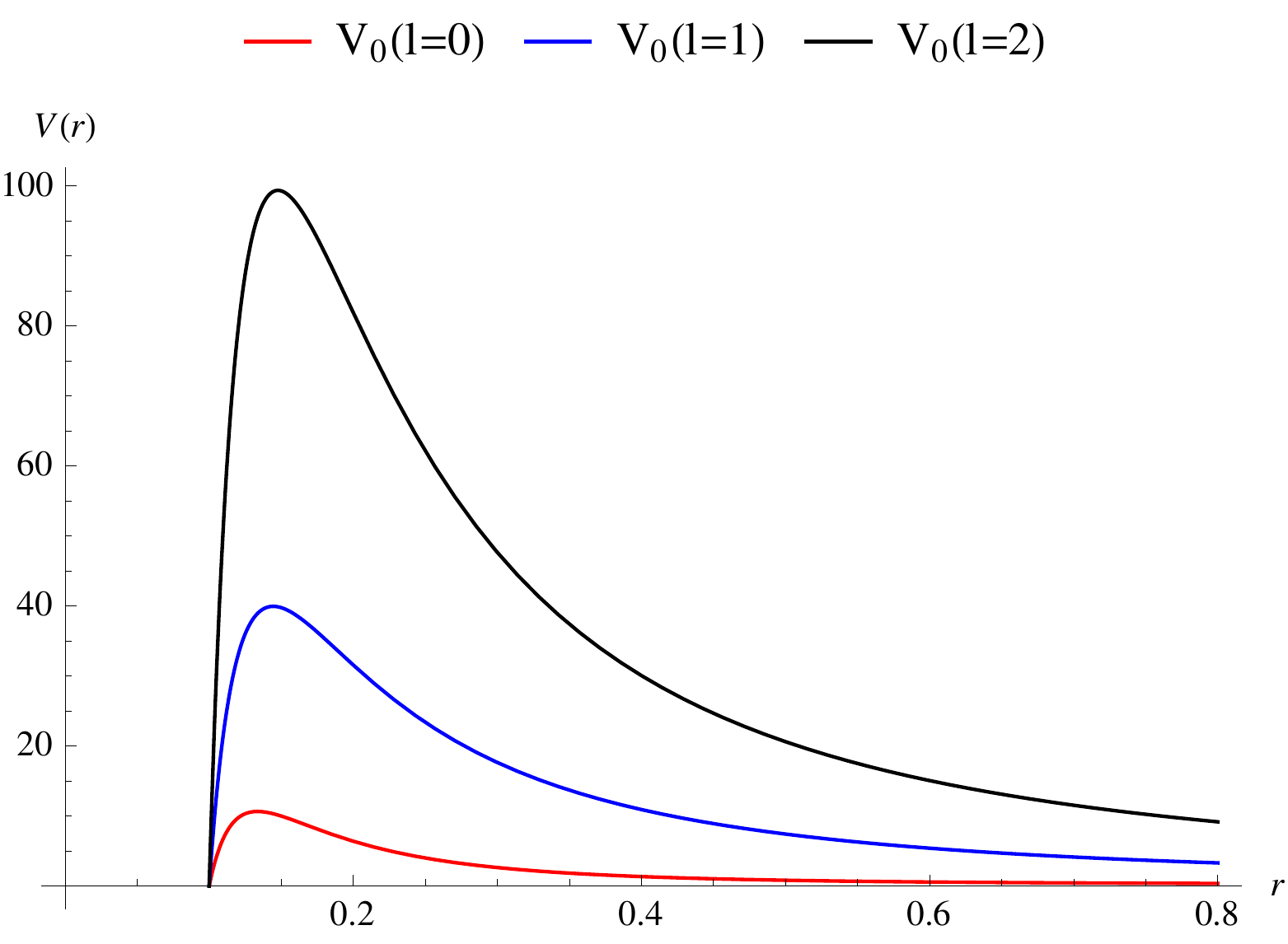} }
\end{minipage}
\hspace{0.29cm}
\begin{minipage}[ht]{0.49\linewidth}
\center{\includegraphics[width=1.\linewidth]{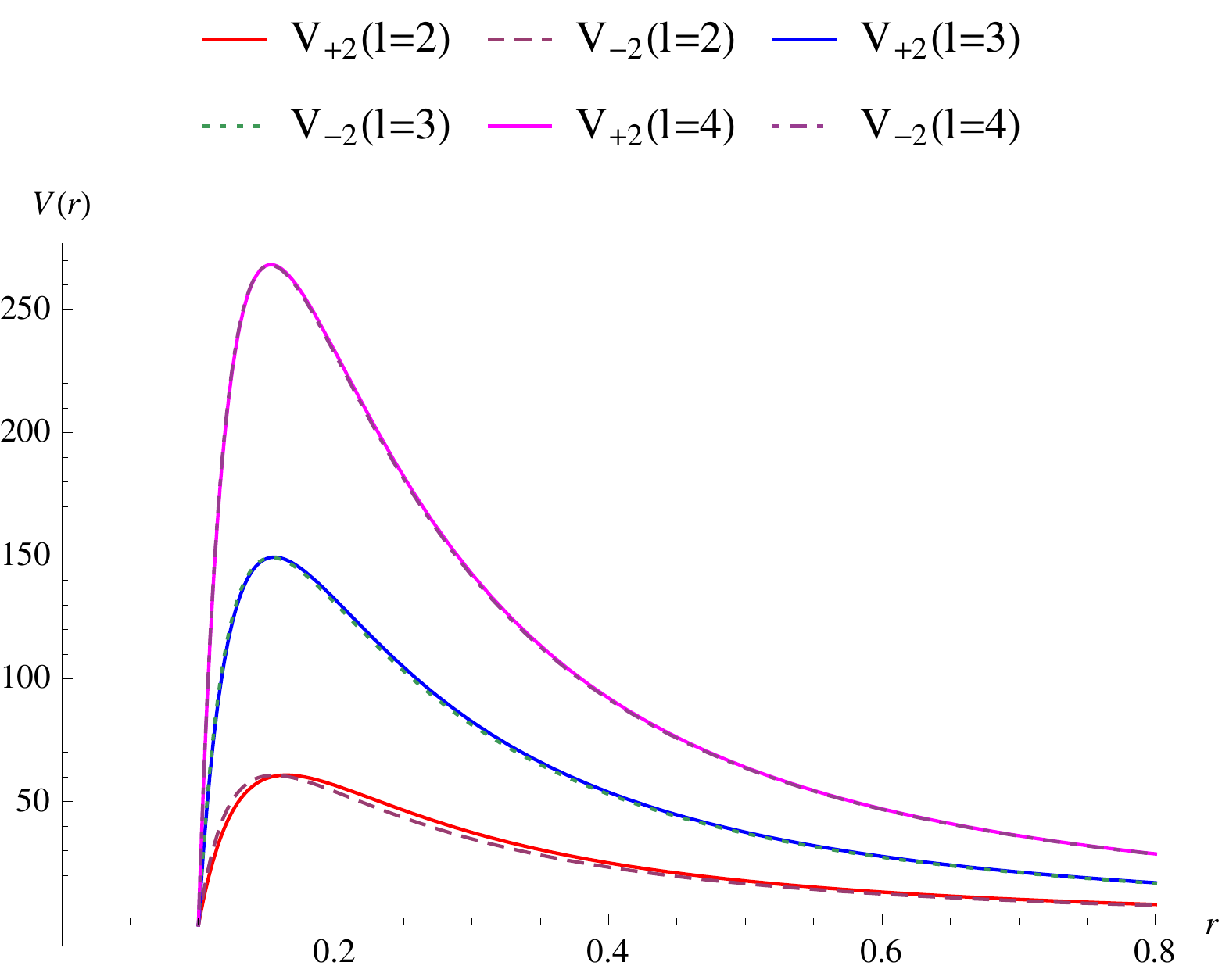}}
\end{minipage}
\caption{\label{fig2} \textbf{Left panel}: the shape of the effective potential of different modes ($l=0,1,2$) of the scalar perturbation over the spherically-symmetric Anti-de Sitter (AdS)-Schwarzschild background near the black hole (BH) horizon. \textbf{Right panel}: comparison of the effective potentials $V_{+2}(r)$ (solid lines) and $V_{-2}(r)$ (dashed lines) of the d-wave perturbations ($l=2,3,4$) over the spherically-symmetric AdS-Schwarzschild background near the BH horizon. The~value of the horizon radial location is $r_+=0.1$. The~AdS cosmological constant is chosen to be $\k^2=-\L/3=0.1$.}
\end{figure}

\section{Generalized Angular Momentum Numbers: Generalities and Axial Symmetry~Case}\label{sec:4}

Let us turn back to the master equation for the fundamental angular variable (\ref{Thetaeq}). It can be solved with series expansion of $\Theta(\theta,\varphi)$ over the spherical harmonics $Y_{lm}$:
\begin{equation}
\Theta(\theta,\varphi)= \sum_{l, m}^\infty c_{lm}Y_{lm}(\theta,\varphi);\quad l\in \mathbb{Z}, \,\,m=-l,\dots,l.
\label{ThetaY}
\end{equation}

For a general function of angles, the expansion (\ref{ThetaY}) contains infinite number of terms. Therefore, once one plugs (\ref{ThetaY}) in the master Equation~(\ref{Thetaeq}), the~solution to the master equation turns into a generalized eigenvalue problem:
\begin{equation}
\sum_{j, m}^\infty A_{km',\; jm}c_{jm} = C\sum_{j, m}^\infty B_{km',\;jm}c_{jm},\quad C=\nu(\nu+1),
\label{eigen}
\end{equation}
with infinite dimensional matrices
\begin{equation}
A_{km',jm} = j(j+1)\delta_{kj}\delta_{m'm},\quad B_{km',jm} = \int d\Omega \,e^{\chi(\theta,\varphi)}\,Y^*_{km'}(\theta,\varphi)Y_{jm}(\theta,\varphi).
\label{ABdef}
\end{equation}

As usual, the measure of integration over the angle variables is determined by $d\Omega=\sin \theta d\theta d\varphi$.

One of the ways to resolve the generalized eigenvalue problem we are dealing with is to use numerical computations. However, even in this case, we have to establish an upper bound for $j$ in (\ref{eigen}) to reduce the task to the generalized eigenvalue problem of $n\times n$ matrices ${\bf A}$ and ${\bf B}$
\be
{\bf A}\cdot {\bf c}=C\,{\bf B}\cdot {\bf c},
\la{genABn}
\ee 
where 
\be
n=\sum_{j=0}^{j_{max}} (2j+1)=(1+j_{max})^2.
\la{ndef}
\ee

The upper bound value $j_{max}$ is chosen in such a way that the eigenvalues and the corresponding eigenvectors do not visibly change upon $j_{max}$ increasing.

Another simplification can be reached with restoring a part of the spherical symmetry, that is the axial symmetry. Viz., the~metric potential $\chi$, generally dependent on two spherical angles, becomes a function of just one polar angle $\theta$. 

In this case, Equation \rf{Lspheq} simplifies to
\be
\frac{1}{\sin\th}\frac{d}{d\th}\left(\sin\th\,\frac{d\chi(\th)}{d\th}\right) + 2\left(e^{\chi(\th)}-1\right) = 0,
\la{LeqAXS}
\ee
so one can establish that (see Appendix A)
\begin{equation}
e^{\chi(x)} = \frac{(2ab)^2\left(\frac{1-x}{1+x}\right)^a}{(1-x^2)\left(b^2 + \left(\frac{1-x}{1+x}\right)^a\right)^2}, \quad x = \cos\theta,
\label{expaxs}
\end{equation}
where the constants $a=1+\alpha$ and $b=1+\beta$ are restricted to be real and positive. (The case of trivial deformation parameters $\alpha$ and $\beta$ corresponds to the spherical symmetry of the background metric.) Additional restriction on the deformation parameters is required by finiteness of the metric potential $e^{\chi(\th)}$ in the endpoints of the $\th$ fundamental domain $\th \in [0,\pi]$. According to this requirement, $\a$ has to be~non-negative.

Axial symmetry makes possible to further separate the angular variables
\begin{equation}
 \Theta\rightarrow\Theta_m(\theta,\varphi) = e^{im\varphi }S_m(\theta), \;\; \Theta_m(\theta,\varphi)  = \Theta_m(\theta,\varphi + 2\pi) \rightarrow m \in \mathbb{Z},
\label{Thetaaxs}
\end{equation}
so that the master Equation~(\ref{Thetaeq}) is reduced to the enlargement of general Legendre equation:
\begin{equation}
\frac{d}{dx}\left[(1-x^2)\frac{dS_m(x)}{dx}\right] + \left[\nu(\nu + 1)e^{\chi(x)} - \frac{m^2}{1-x^2}\right]S_m(x) = 0, \;\; x = \cos{\theta}.
\label{Stheq}
\end{equation}

Consequently, the~matrices (\ref{ABdef}) are reduced to
\begin{equation}
A^m_{ij} = j(j+1)\frac{2(j+m)!}{(2j+1)(j-m)!}\delta_{ij}, \quad B^m_{ij} = \int_{-1}^{1}dx \,e^{\chi(x)}P_{im}(x)P_{jm}(x),
\label{ABaxs}
\end{equation}
and the generalized eigenvalue problem \rf{eigen} turns into
\be
\sum_{j=|m|}^\infty A^m_{ij} c_j=C \sum_{j=|m|}^\infty B^m_{ij} c_j.
\la{geneigenax}
\ee

Solving for Equation~(\ref{geneigenax}), we arrive at the following conclusions: the eigenvalues $\nu$'s are labeled with two indices---$l\ge 0$ and $m=\{-l,\dots,l\}$---with the trivialization condition $\nu_{lm} \rightarrow l \in \mathbb{Z}$ once $\chi\rightarrow 0$; numerics also give $\nu_{l0} = l$ and $\nu_{l,-m}=\nu_{lm}$. Computations of the separation constant $C_{lm}=\n_{lm}(\n_{lm}+1)$ for different values of the admissible deformation degrees $\a$ and $\b$ show the independence of the results from the value of $\b$. See Table~\ref{tab1} as an example. Therefore, $\n_{lm}$ do not depend on $\b$ and are solely functions of the parameter~$\a$.

\begin{table}[ht]
\captionsetup{font=small}
\caption{\label{tab1} Values of $C_{lm} = \n_{lm}(\n_{lm}+1)$ with $l = 1, m = 1$ for different deformation degrees $\a$ and $\b$ obtained by solving for eq. \rf{geneigenax} with different cut-off values of $j_{max}$.}
\centering
\begin{tabular}{cccc}
\hline\hline
$\boldsymbol{\a}$ & $\boldsymbol{\b}$ & $\boldsymbol{j_{max}=42}$ & $\boldsymbol{j_{max}=72}$\\
\hline
$\a = 0.001$ & $\b = 0.001$ & 1.99873&1.99700\\
$\a = 0.001$ & $\b = 0.01$ &1.99873&1.99700\\
 $\a = 0.001$ & $\b = 0.1$ &1.99873 &1.99700\\
 $\a = 0.001$ & $\b = 1.0$ & 1.99873 &1.99700\\
\hline
$\a = 0.01$ & $\b = 0.001$ &1.97164&1.97039\\
$\a = 0.01$ & $\b = 0.01$ &1.97164&1.97039\\
$\a = 0.01$ & $\b = 0.1$ & 1.97164 &1.97039\\
$\a = 0.01$ & $\b = 1.0$ & 1.97164 &1.97039\\
\hline
$\a = 0.1$ & $\b = 0.001$ &1.73554 &1.73553\\
$\a = 0.1$ & $\b = 0.01$ &1.73554 &1.73553\\
$\a = 0.1$ & $\b = 0.1$ &1.73554 &1.73553\\
$\a = 0.1$ & $\b =1$ &1.73554 &1.73553\\
\hline
$\a = 1.0$ & $\b = 0.001$ & 0.74999 &0.74999\\
$\a = 1.0$ & $\b = 0.01$ & 0.74999 &0.74999\\
$\a = 1.0$ & $\b = 0.1$ & 0.74999 &0.74999\\
$\a = 1.0$ & $\b = 1$ & 0.74999 &0.74999\\
\hline\hline
\end{tabular}
\end{table}

Next, fixing $b = 1$ and $a = 1 + \alpha$,
we observe from numerics that first eigenvalues with $l,m = 1, 2$ as functions of a positive parameter $\alpha$ obey the inequality $\nu_{lm} \leq l$ (see Figure~\ref{figure:fig-3}). This trend gets preserved for other combinations of $(l,m)$ in $\n_{lm}$, as well. (Cf. Table~\ref{tab2}.)
\begin{figure}[ht]
\begin{center}
\includegraphics[width=6.6cm]{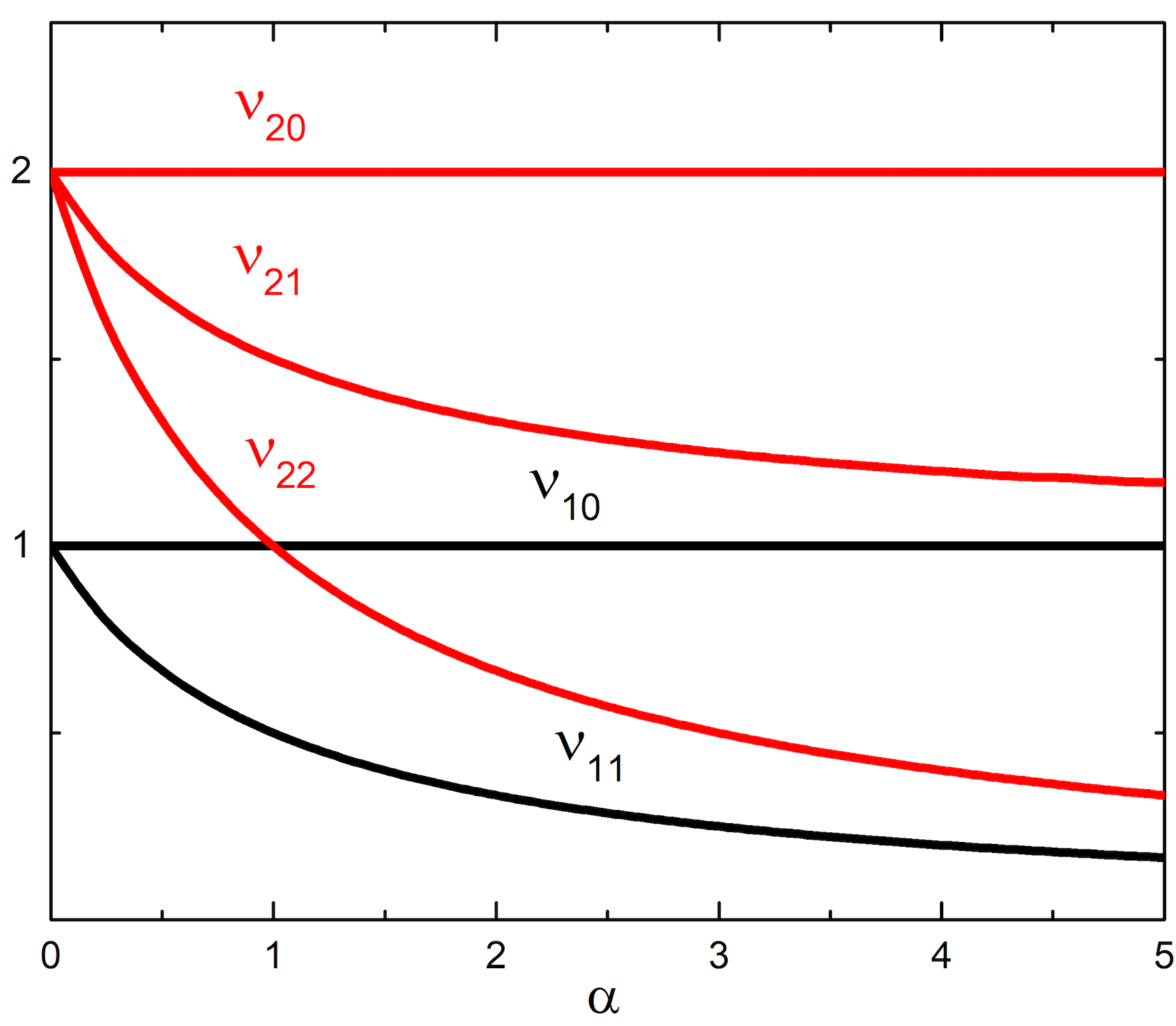}
\end{center}
\caption{Egenvalues $\nu_{lm}$ for $l=1,2$ and $m\le l$.}
\la{figure:fig-3}
\end{figure}

Finally, comparing different values of $\nu_{lm}$ for different values of the deformation degree $\alpha$, we observe that (see Table~\ref{tab2})
\begin{equation}
\nu_{lm} = \nu_{l-1,m} + 1.
\label{nuquant}
\end{equation}

Therefore, we conclude that the angular momentum is quantized, but~not in~integers.

\begin{table}[ht]
\captionsetup{font=small}
\caption{\label{tab2} Values of $\n_{lm}$ ($l=0,\dots,5$, $m=0,\dots,4$) for different values of the deformation parameter $\a$.}
\centering
\begin{tabular}{ c c c c c c c}
\hline\hline
  & $\boldsymbol{\a=0.001}$ & $\boldsymbol{\a=0.01}$ & $\boldsymbol{\a=0.1}$ & $\boldsymbol{\a=1}$ & $\boldsymbol{\a=2}$ & $\boldsymbol{\a=3}$ \\ 
\hline
$\n_{00}$ & 0 & 0  & 0 & 0 & 0 & 0\\  
$ \n_{10}$ & 1 & 1   & 1 & 1 & 1 & 1\\
$\n_{20}$ & 2 & 2 & 2 & 2 & 2 & 2\\
$\n_{30}$ & 3 & 3 & 3 & 3 & 3 & 3\\
$\n_{40}$ &4 & 4& 4& 4 & 4 & 4 \\
$\n_{50}$ &5 &5 &5 &5 & 5 & 5 \\
\hline
$\n_{11}$ &0.999 &0.990 &0.909 &0.5 & 0.333&0.25\\
$\n_{21}$ &1.999 &1.990 &1.909 &1.5 & 1.333&1.25\\
$\n_{31}$ &2.999 &2.990 & 2.909&2.5 & 2.333&2.25\\
$\n_{41}$ &3.999 &3.990 &3.909 &3.5 & 3.333&3.25\\
$\n_{51}$ &4.999 &4.990 & 4.909&4.5 & 4.333&4.25\\
\hline
$\n_{22}$ &1.9980 &1.9802 &1.8181 &1 &0.666&0.5\\
$\n_{32}$ &2.9980 &2.9802 &2.818 &2 &1.666&1.5\\
$\n_{42}$ &3.9980 &3.9802 &3.818 &3 &2.666&2.5\\
$\n_{52}$ &4.9980 &4.9802 & 4.818&4 &3.666&3.5\\
\hline
$\n_{33}$ &2.9970 &2.9702 &2.727 &1.5 &1.0&0.75\\
$\n_{43}$ &3.9970 &3.9702 &3.727 &2.5 &2.0&1.75\\
$\n_{53}$ &4.9970 &4.9702 &4.727 &3.5 &3.0&2.75\\
\hline
$\n_{44}$ &3.9960 &3.9604 &3.6363 &2 &1.333&1.0\\
$\n_{54}$&4.9960 &4.9604 & 4.6363&3 &2.333&2.0\\
\hline\hline
\end{tabular}
\la{table:table1}
\end{table}

We end up this section with an analysis of Table~\ref{tab2}, which results in the following analytical expression for $\n_{lm}(\a)$:
\be
\n_{lm}(\a)=l-\fr{\a}{1+\a}\,m.
\la{nualpha}
\ee

The obtained expression is in the fine agreement with the previously numerically computed functional dependences of $\n_{lm}$ on $\a$ (Figure \ref{figure:fig-3}). Let us also notice the coincidence in $\n_{lm}=\nu_{l+k,m+s}$ values for integers $(k,s)$, determined by $k(1+\a)=\a s$ and restricted by $l+k\ge m+s$. For~instance, $\n_{00}=\n_{33}$, $\n_{20}=\n_{43}=\n_{66}$, $\n_{21}=\n_{44}$, $\n_{31}=\n_{54}=\n_{77}$, and so~on.

\section{The Grey-Body Factor: Schwarzschild vs Distorted \\Schwarzschild}\label{sec:5}

Now, let us turn to the radial part of the d-wave perturbations described by \mbox{Equation~(\ref{RWZeqs})}. Solving for these equations analytically is hampered due to a complicated form of the effective potentials. Typically, one has to find solutions either numerically (as it was done in ref.~\cite{Gray:2015xig}), or~to follow the procedure of finding solutions in different coordinate domains---the near, mid and far zones (see, e.g., ref.~\cite{Harmark:2007jy}, for a review)---with subsequent constructing the united solution in different approximations (as, for~instance, in refs.~\cite{Starobinskil:1974nkd,Unruh:1976fm,Sanchez:1976fcl,Sanchez:1976xm,Sanchez:1977si,Sanchez:1977vz}). In~the context of the scattering problem, solutions to Equation~(\ref{RWZeqs}) are used in computing different cross-sections, one of important ingredients of which is the so-called grey-body factor (GBF). To~compute this characteristic, one has to solve a Schr\"odinger-like \mbox{Equation~(\ref{RWZeqs})} with the specified boundary conditions (see, for~example, ref.~\cite{Harmark:2007jy}), which especially determine the transmission coefficient of an ingoing wave through the barrier of effective potentials (\ref{V+2}) or (\ref{V-2}). Then, the~grey-body factor is $\gamma(\omega)=|T(\omega)|^2$, where $T(\omega)$ stands for the transmission coefficient. The~complete transmission means the complete absorption of incoming waves by a~BH.

To figure out hallmarks of scattering/absorption in the background of quasi-spherical/distorted BHs, we will compare the spherically symmetric case (here, we follow the approach of ref.~\cite{Gray:2015xig}) with that of deformed but axially symmetric. Looking at Figure~\ref{fig4} with the results of numerics for the spherically symmetric (Schwarzschild) background (which reproduce in part data in Figures~2 and 8 of ref.~\cite{Gray:2015xig}), one may notice~that:
\begin{itemize}
\item
the scalar $s$-wave (left panel) has the complete transmission at the lowest admissible value of frequency $\omega$;
\item
increasing $l$ in the scalar mode perturbations (p- and d-modes on the left panel) requires higher values of $\omega$ to reach the complete transmission;
\item
basic axial gravitational perturbations ($d$-waves) (right panel) reach the complete transmission at lower, w.r.t. $l=2$ mode of axial electro-magnetic (EM) and scalar perturbations, frequency.
\end{itemize}

\begin{figure}[h]
\begin{minipage}[ht]{0.49\linewidth}
\center{\includegraphics[width=0.98\linewidth]{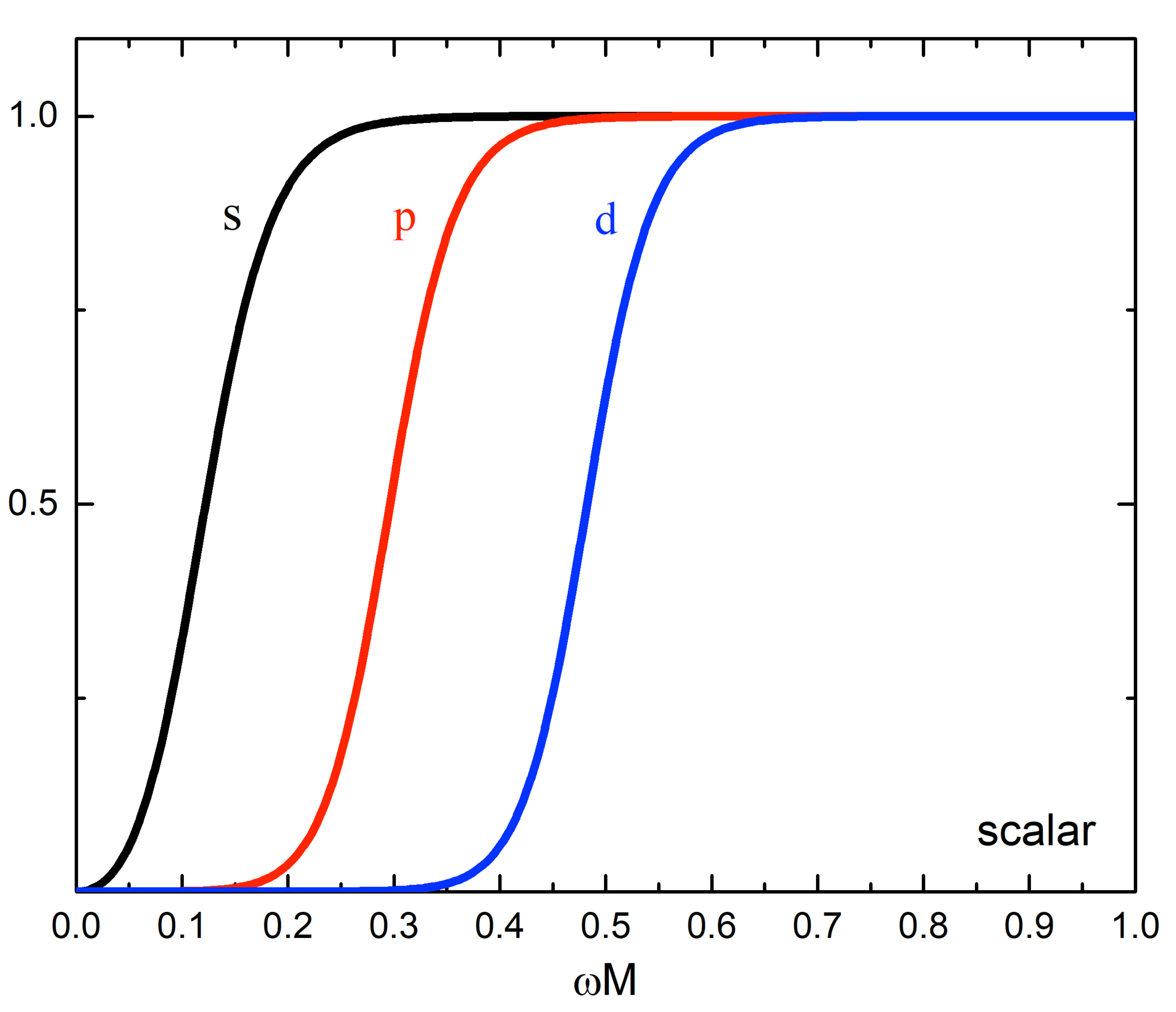} }
\end{minipage}
\hspace{0.29cm}
\begin{minipage}[ht]{0.49\linewidth}
\center{\includegraphics[width=1\linewidth]{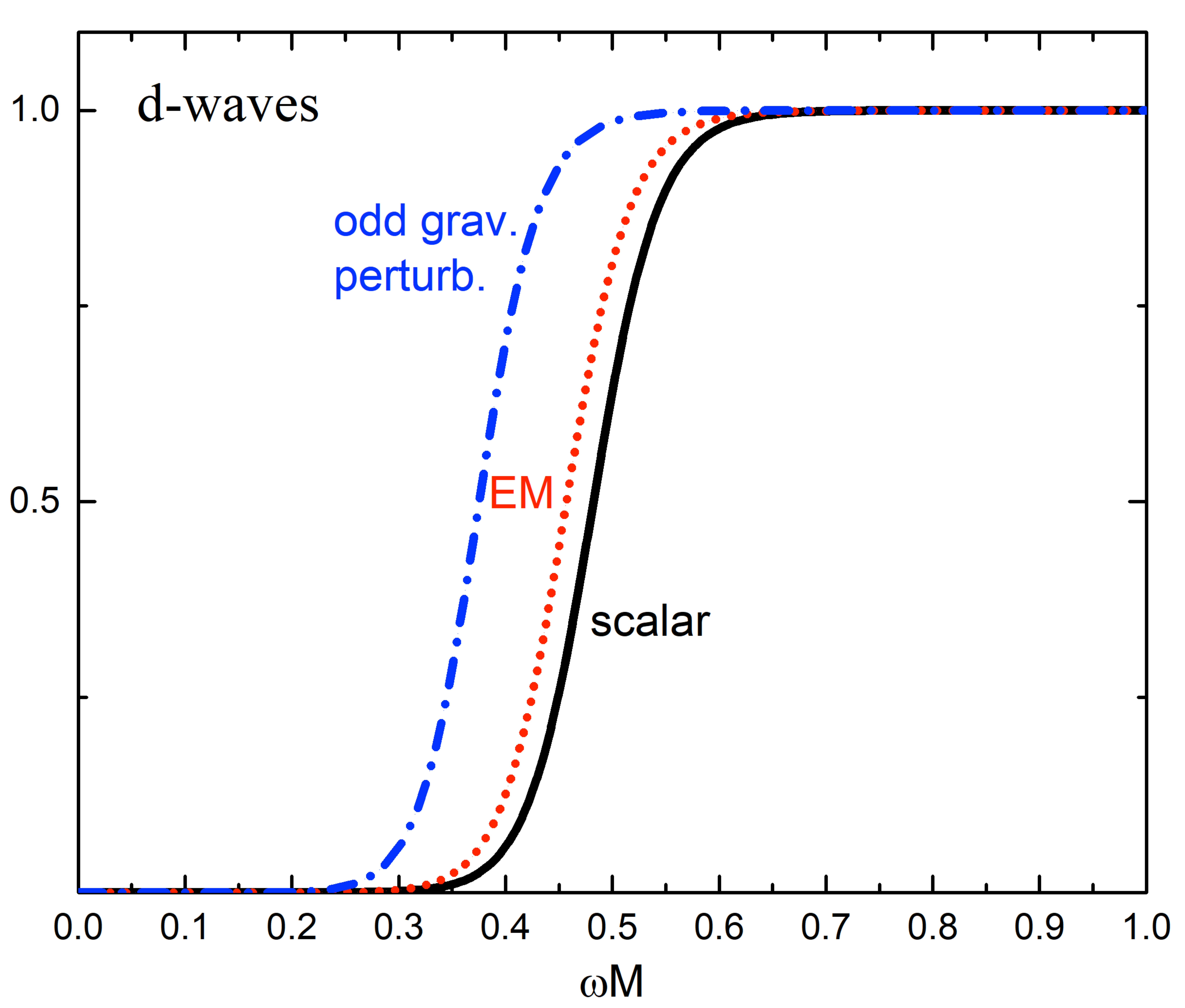}}
\end{minipage}
\caption{\label{fig4} \textbf{Left panel}: the grey-body factors
$\gamma(\omega)$ of different modes ($l=0,1,2$) of the scalar perturbation over the Schwarzschild background. \textbf{Right panel}: comparison of $\gamma(\omega)$ of the basic axial d-wave perturbation ($l=2$) to the corresponding mode (with $l=2$) grey-body factors (GBFs) of the scalar and axial {EM} 
perturbations over the Schwarzschild~background.}
\end{figure} 

For the distorted BH background (\ref{ds2def}) with the metric potential (\ref{expaxs}), the~deformation parameters of which are chosen to be $b=1$ and $a=1+\alpha=1+0.2$, putting the data for the same type of waves on plots, we encounter important differences in compare to the previously considered cases (see Figure~\ref{fig5}). We observe~that:
\begin{itemize}
\item
for each value of $l$ (recall, $l \in \mathbb{Z}$ is a non-negative degree of the corresponding spherical harmonics in the series expansion (\ref{ThetaY})), there are $l+1$ different values of the grey-body factor $\gamma(\omega)$;
\item
$\gamma(\omega)$ gets increased with increasing the deformation degree $\alpha$;
\item
for a fixed $l$, the~GBFs with the maximal projection value $m = l$ reach the complete transmission at the lowest values of $\omega$.
\end{itemize}
\begin{figure}[ht]
\begin{minipage}[ht]{0.49\linewidth}
\center{\includegraphics[width=0.98\linewidth]{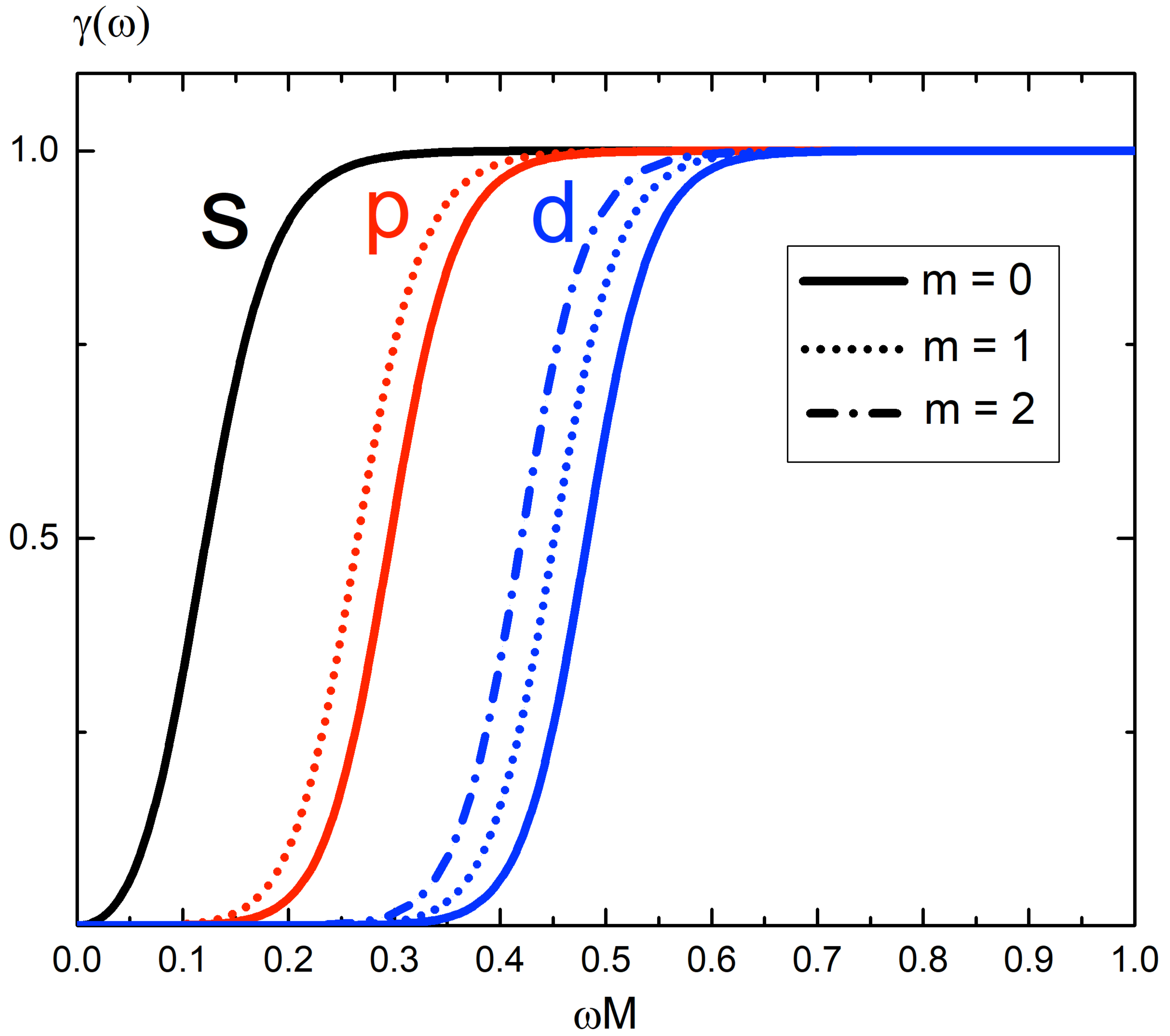} }
\end{minipage}
\hspace{0.29cm}
\begin{minipage}[ht]{0.49\linewidth}
\center{\includegraphics[width=1\linewidth]{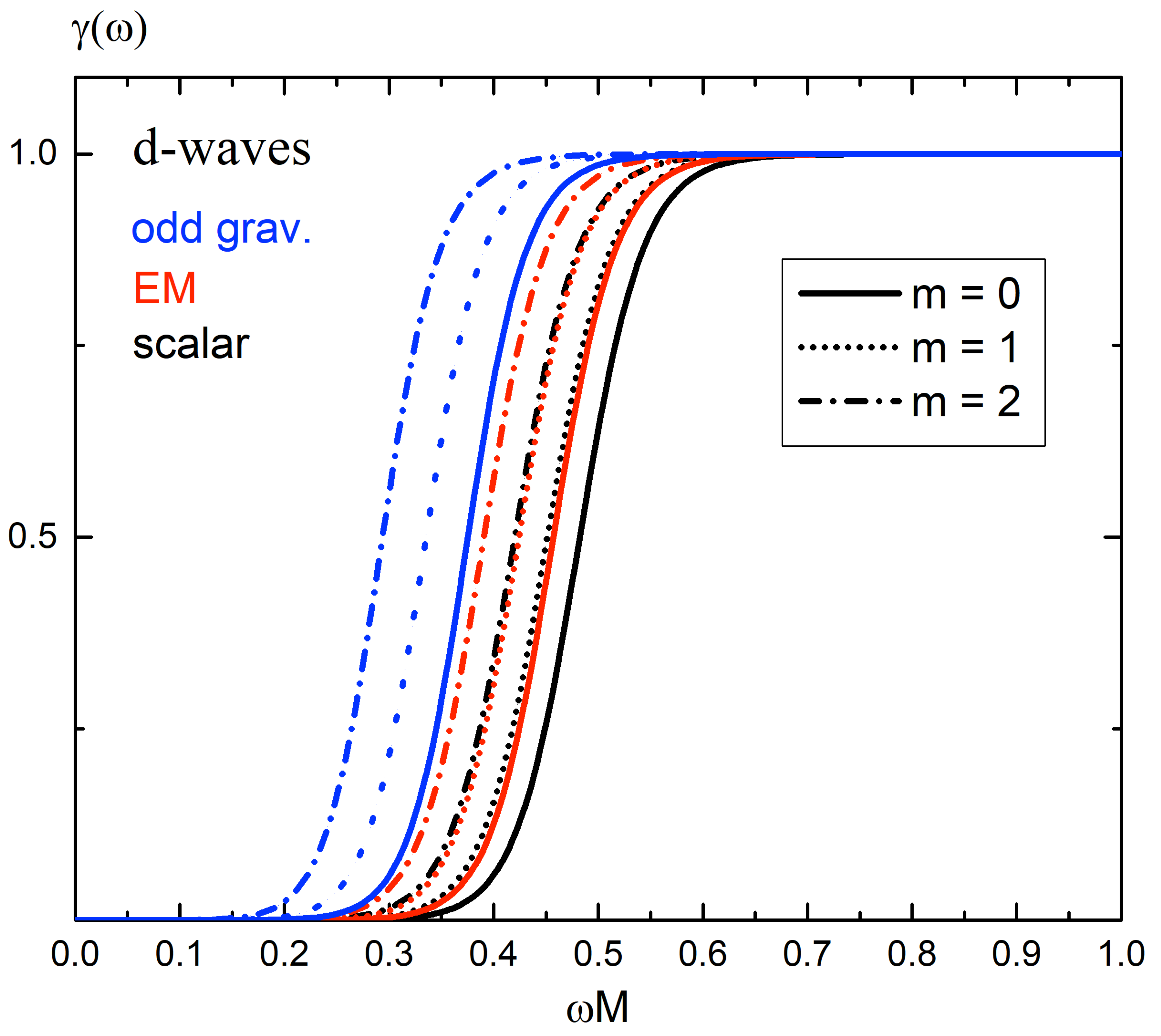}}
\end{minipage}
\caption{\label{fig5} \textbf{Left  panel}: splitting the grey-body
factors $\gamma(\omega)$ for different modes ($l=0,1,2$) of the scalar perturbation in the distorted BH background. \textbf{Right panel}: comparison of the GBFs of the basic axial d-wave perturbation ($l=2$) to that of the scalar and axial EM perturbations over the distorted BH background. The~deformation parameter $\alpha$ is equal to $0.2$.}
\end{figure}

\section{Quasinormal Modes of a Quasi-Spherical Axisymmetric Black~Hole }\label{sec:6}

As we have noted above, the~scattering problem for an axially-symmetric neutral BH is completely determined by a single parameter---the deformation degree $\a$---which is required to be non-negative without any additional limitations. However, new restrictions on $\a$ could appear from the demand on stability of the BH spacetime background against small perturbations. That is fully determined by the quasinormal modes (QNMs). 

Recall that the~quasinormal modes (see, e.g., refs.~\cite{Chandr,Otsuki91,Horowitz:1999jd,Cardoso:2001bb,Cardoso:2003cj,Konoplya:2011qq,Moulin:2019bfh,Arslanaliev:2018ked,Konoplya:2003ii,Lin:2016sch,Leaver:1985ax,Nollert:1999ji,Berti:2003ud,Ferrari:2007dd,Matyjasek:2017psv,Konoplya:2019hlu}) correspond to solutions to Equations \rf{RWZeqs} and (\ref{s012axeqs}), which satisfy the specific boundary conditions (cf.~\cite{Chandr,Horowitz:1999jd,Cardoso:2001bb}): \\ingoing waves at the horizon
\be
Q_s(x) \simeq e^{-i\w r_*}, \;\;\; r_*\rightarrow-\infty\; (r\rightarrow r_+),
\la{bc_1_QNM}
\ee
and outgoing waves at the spatial infinity 
\be
\begin{split}
&Q_s(x) \simeq e^{i\w r_*}, \;\; r_*, r \rightarrow +\infty \;\;\;\text{flat space},\\
&Q_s(x) \rightarrow 0,\; \;\;\;\; r_*, r \rightarrow +\infty\;\;\;\text{AdS space}\;.
\end{split}
\la{bc_2_QNM}
\ee

For fixed values of spin $s$ and (projection of) angular momentum $(l,m)$, there are infinite number of QNMs, which are labeled by the overtone number $n$; the least damped (fundamental) QNMs correspond to $n=0$.

Generally, the~QNMs are complex and their imaginary part can be positive or negative. It is easy to verify~\cite{Horowitz:1999jd,Cardoso:2001bb} that positivity in the imaginary part of any QNM results in the instability of the background geometry. Indeed, following refs.~\cite{Horowitz:1999jd,Cardoso:2001bb}, we will find solutions to the Schr\"odinger-like equation
\begin{equation}
\left[\frac{\partial^2}{\partial r_*^2}  + \omega^2 - V_s(r)\right] Q_{s} = 0,  \;\;r_* \in(-\infty, +\infty), 
\label{RWZeqs1}
\end{equation}
with the ``wave-function'' 
\be
Q_s(r)=e^{-i\w r_*}\f_s(r),
\la{Qsans}
\ee
and figure out restrictions on the admissible complex frequencies. To~maintain the boundary conditions \rf{bc_1_QNM}, \rf{bc_2_QNM}, the~``amplitude'' $\f_s(r)$ should satisfy
\vspace{6pt}
\be
\f_s(r) \simeq \mathrm{const},\;\;\;\; r\ra r_+,
\la{fbc1}
\ee
and
\be
\begin{split}
&\f_s(r) \ra e^{2i\w r_*}, \;\; r_*, r \rightarrow +\infty \;\;\;\text{flat space},\\
&\f_s(r) \rightarrow 0,\;\;\; \;\;\;\; r_*, r \rightarrow +\infty\;\;\;\text{AdS space}\;.
\end{split}
\la{fbc2}
\ee

Plugging the ansatz \rf{Qsans} into Equation \rf{RWZeqs1}, for~inherently complex valued $\f_s(r)$, we~get
\be
f\frac{d^2\phi_s}{dr^2} +\left(\fr{df}{dr}- 2i\w\right)\frac{d\phi_s}{dr} - \frac{V_s}{f}\phi_s = 0,\quad f=1-\fr{2M}{r}+\k^2 r^2.
\la{feq}
\ee

Now, one multiplies both sides of \rf{feq} by $\phi_s^*(r)$ and integrates over $[r_+,\infty)$: 
\be
\int_{r_+}^{\infty} dr\left(\phi_s^*\frac{d(f\pa_r\phi_s)}{dr}-2i\w\phi_s^*\frac{d\phi_s}{dr}-\frac{V_s}{f}|\phi_s|^2\right) = 0\;.
\la{int1}
\ee

Further integration by parts on account of the boundary conditions \rf{fbc1}, \rf{fbc2} and the asymptotic behaviour of the red-shift factor $f(r)$ at the integration end points results in the following expression for the first term of the integrand in \rf{int1}:
\be
\int_{r_+}^{\infty} dr\frac{d(\phi_s^*f\pa_r \phi_s)}{dr} = \phi_s^*f\frac{d\phi_s}{dr}\bigg|_{r = r_+}^{r = \infty} = \begin{cases}&2i\w\;\;\;\text{flat}\\
	&0\,\,\,\;\;\;\;\text{AdS}\end{cases}.
\ee

Hence,
\be
\int_{r_+}^{\infty} dr\left( f\bigg|\frac{d\phi_s}{dr}\bigg|^2+2i\w\phi_s^*\frac{d\phi_s}{dr}+\frac{V_s}{f}|\phi_s|^2\right) = \begin{cases}&2i\w\;\;\;\text{flat}\\
	&0\,\,\,\;\;\;\;\text{AdS}\end{cases}\;,
\la{cond}
\ee
and, for the imaginary part of \rf{cond}, we obtain
\be
\int_{r_+}^\infty dr \left(\w\phi_s^*\frac{d\phi_s}{dr} + \w^*\phi_s\frac{d\phi_s^*}{dr}\right)  = \begin{cases}&\w + \w^*\;\;\;\text{flat}\\
	&0\,\,\,\;\,\,\,\;\,\,\,\;\;\;\;\;\text{AdS}\end{cases}\;.
\la{cond1}	
\ee

Using the integration by parts for the left-hand-side (l.h.s.) 
of \rf{cond1} once again results in
\be
\int_{r_+}^\infty dr \left(\w\phi_s^*\frac{d\phi_s}{dr} + \w^*\phi_s\frac{d\phi_s^*}{dr}\right)  = (\w - \w^*)\int_{r_+}^\infty \phi_s^*\frac{d\phi_s}{dr}dr + \w^*|\phi_s(r)|^2\bigg|_{r=r_+}^{r=\infty} ,
\ee
so that
\be
\int_{r_+}^\infty \phi_s^*\frac{d\phi_s}{dr}dr  = \begin{cases}&\frac{\w + \w^*|\phi_s(r_+)|^2}{\w-\w^*}\;\;\;\text{flat}\\
	&\frac{\w^*|\phi_s(r_+)|^2}{\w-\w^*}\,\,\,\;\;\;\;\;\text{AdS}\end{cases}\;.
\la{cond2}
\ee

Lastly, substituting \rf{cond2} into the l.h.s. of \rf{cond} leads to~\cite{Horowitz:1999jd,Cardoso:2001bb}
\be
\int_{r_+}^{\infty} dr\left( f\bigg|\frac{d\phi_s}{dr}\bigg|^2+\frac{V_s}{f}|\phi_s|^2\right) = \begin{cases}&-\frac{|\w|^2|\phi_s(r_+)|^2 +(\text{Re}\,\w)^2 + (\text{Im}\,\w)^2}{\text{Im}\,\w}\;\;\;\text{flat}\\
	&-\frac{|\w|^2|\phi_s(r_+)|^2}{\text{Im}\,\w}\,\,\,\;\,\,\,\;\,\,\,\;\,\,\,\;\,\,\,\;\,\,\,\;\,\,\,\;\,\,\,\;\;\;\text{AdS}\end{cases}\;.
\la{cond3}
\ee

Therefore, for~a non-negative in the domain $[r_+,\infty)$ potential $V_s(r)$ (cf. Figures~\ref{fig1} and~\ref{fig2} with effective potentials in spherically-symmetric spacetimes), the l.h.s. of \rf{cond3} is non-negative. The~same is required for the right-hand-side (r.h.s.) 
of this expression that means negativity of the denominator: $\text{Im}\,\w < 0$. Once the imaginary part of frequencies becomes positive, it corresponds to an unphysical solution because~fields begin exponentially growth at spatial infinity and near the horizon. Put differently, this situation occurs when the effective potential $V_s(r)$ turns out to be negative within the physical domain $[r_+,\infty)$.

Below, we will explore a possibility to find negative branches of the effective potentials for small perturbations over the quasi-spherical neutral BH background with axial symmetry in Minkowski and AdS~spacetimes.

\subsection{Flat~Spacetime}

In Minkowski spacetime with $f(r) = 1 - r_+/r$, the effective potential (see \rf{Veffax} for $\k^2=0$)
\be
V_s = \frac{f(r)}{r^2}\left((1-s^2)\frac{r_+}{r} + \n(\n+1)\right)
\ee
could be negative only for odd (axial) gravitational perturbations with $s = +2$ (clearly, the~polar tensor perturbations with the Zerilli potential \rf{V-2} do not satisfy this condition); it happens for
\be
\n(\n+1)< \frac{3r_+}{r}, \;\; r\in [r_+,\infty).
\la{V2neg}
\ee

For $\n$'s satisfying \rf{V2neg}, $V_{+2}(r)$ is negative within the interval $r\in[r_+,r_0)$ with $r_0 = \frac{3r_+}{\n(\n+1)}$. Then, from~$r_0>r_+$, we get the following condition on the separation constant: $\n(\n+1)<3$, or~$\n<1.303$. {\it Eigenvalues $\n_{2m}$ smaller than the critical value $\n_{cr}=1.303$ produce the negative effective potential.} Recall that, in the spherically symmetric case, $\n \rightarrow l  \ge 2$, so that the separation constant lowest value is $l(l+1) = 6$. Hence, the~effective potential $V_s(r)$ for $s=0,1,\pm 2$ is always positive. According to \rf{cond2} that gives $\text{Im}\,\w <0$ and the standard, Schwarzschild background is stable against small perturbations~\cite{Regge:1957td,Zerilli:1970se,Chandr}.

In contrast, small perturbations over a quasi-spherical axially-symmetric BH background are characterized by three different values of $\n_{lm}$ with $l=2$, viz., $\n_{20},\n_{21},\n_{22}$. Their dependences on the deformation degree $\a$ are depicted in Figure~\ref{figure:fig-3} in accordance to the relation \rf{nualpha}. One can notice that there are values of $\a$ for which $\n_{22}$ and $\n_{21}$ become smaller of the critical value $\n_{cr}=1.303$; hence, we can expect the appearance of QNMs with a positive imaginary~part.

To find the QNMs, we follow the semianalytical Pad\'e approximation~\cite{Konoplya:2003ii,Matyjasek:2017psv,Konoplya:2019hlu}, improving the standard {Wentzel-Kramers-Brillouin} (WKB) 
technique, and~the numerical Leaver method~\cite{Leaver:1985ax} based on the continued fraction. As~we have discussed, we are mainly interested in values of the QNMs frequencies corresponding to tensor perturbations with eigenvalues $\n_{20},\n_{21},\n_{22}$. The~eigenvalue $\n_{20}$ coincides with that of the spherically symmetric case; therefore, all possible QNMs (fundamental and overtones) related to this case will have the negative imaginary part. The~other QNMs for eigenvalues $\n_{21},\n_{22}$ (of frequency $\w_{21}$ and $\w_{22}$, respectively) are functions of the deformation degree $\a$. The~corresponding frequencies (actually, their real and imaginary parts) are shown in Figure~\ref{fig6}. We find that, for $\a>1$, the imaginary part of $\w_{22}$ becomes greater than zero (see Figure~\ref{fig7}). The~critical value of the deformation parameter $\a=1$ corresponds to $\nu_{22}= 1$ (cf. \rf{nualpha}), so that the domain of negativity of $V_{+2}(r)$ is determined by $[r_+, \frac{3}2r_+)$. Increasing $\a$, this domain increases; $\text{Im}\,\w_{22}$ will increase, too. For~some values of $\a>1$, the imaginary parts of other tensor QNMs also become positive. To~sum up, above~the critical value of the deformation degree $\a=1$, the spacetime geometry of a distorted/quasi-spherical BH becomes unstable against small tensor~perturbations.

\begin{figure}[ht!]
\centering	
	\includegraphics[width=15.6cm]{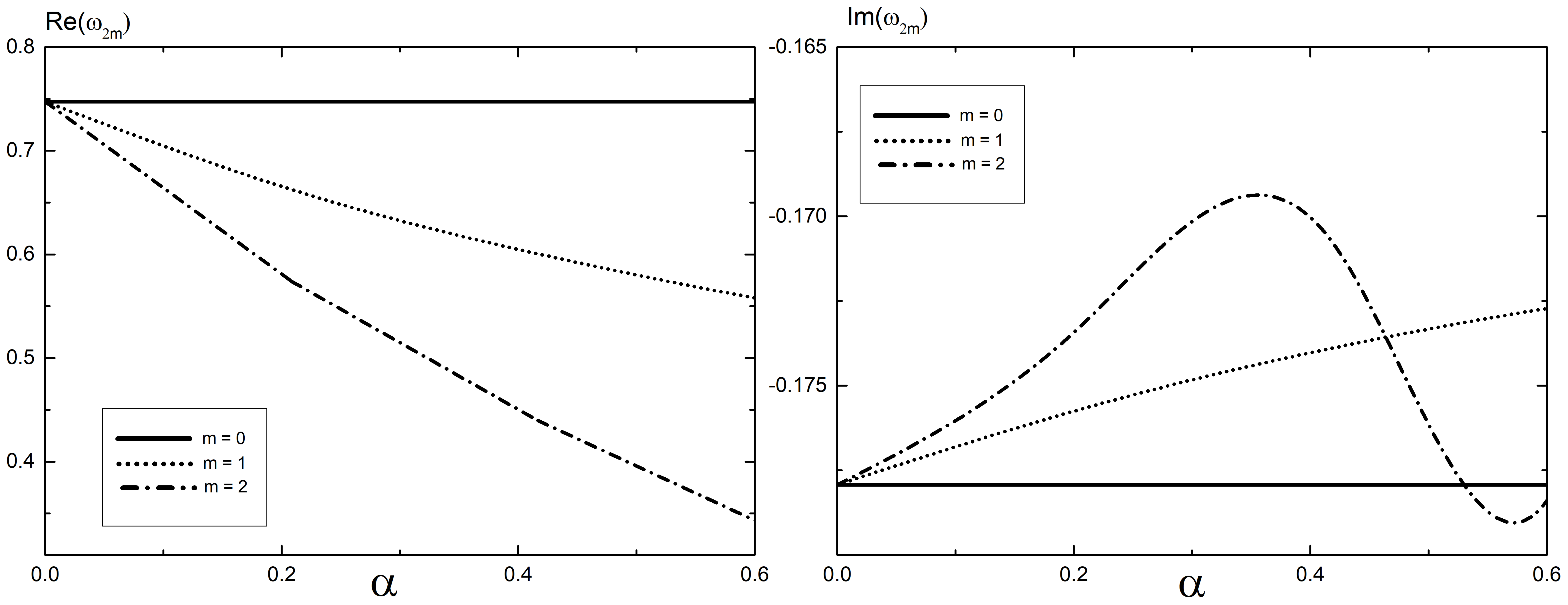}
	\caption{\label{fig6} \textbf{Left panel}: the real part of the
	lowest odd gravitational perturbations $\w_{2m}(\a)$ over a quasi-spherical axially-symmetric neutral BH background in flat~spacetime. \textbf{Right panel}: the imaginary part of $\w_{2m}(\a)$ under the same conditions.} 
\end{figure}
\begin{figure}[ht!]
\centering
	\includegraphics[width=7cm]{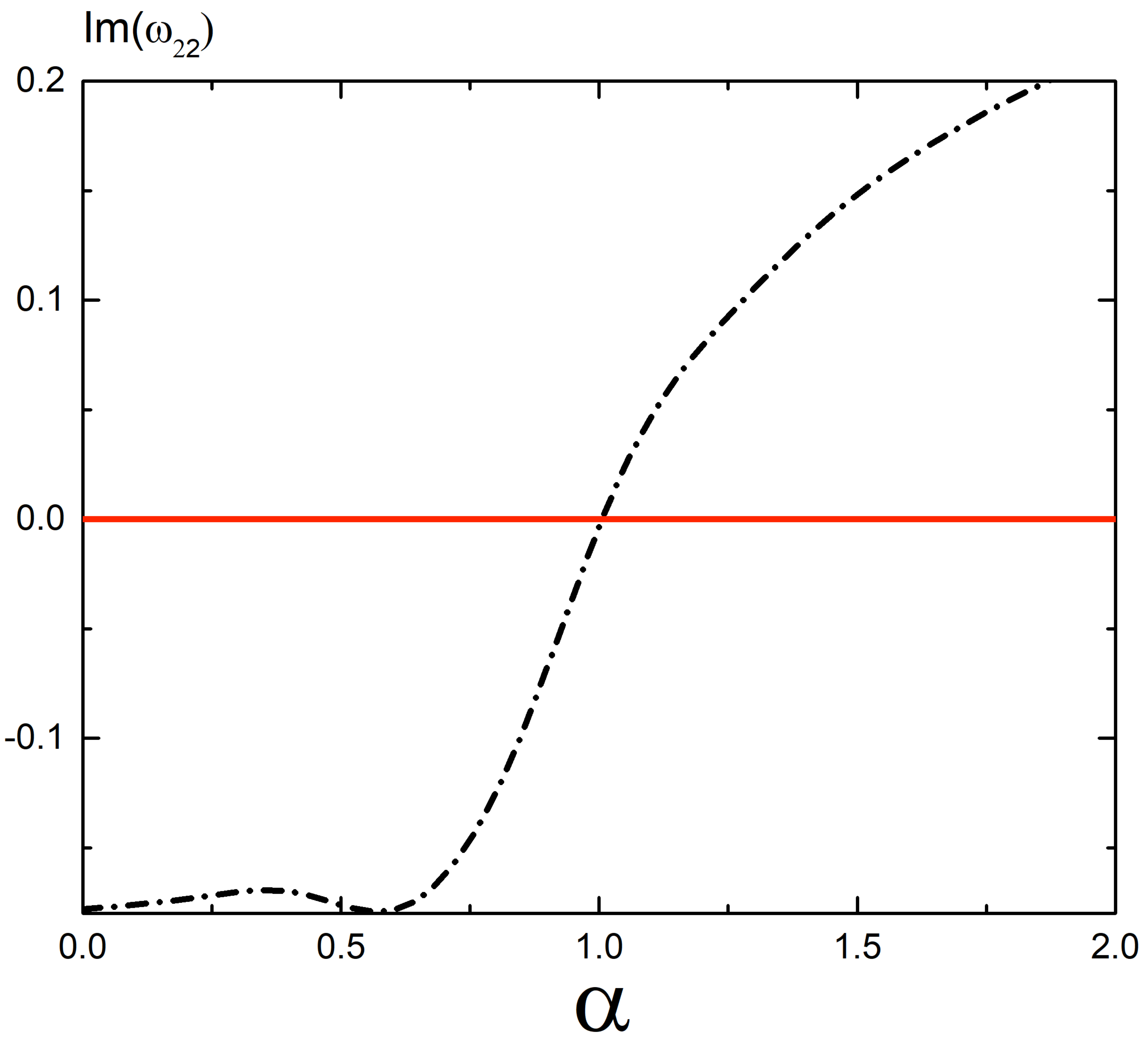}
	\caption{\label{fig7} The imaginary part of $\w_{22}(\a)$ in more detail.} 
\end{figure}

\subsection{AdS~Spacetime}

Now, turn to AdS spacetime. Here, we have $f(r) = 1 + r^2\k^2 - \frac{r_+}{r}(1+  \k^2r_+^2)$, and the effective potential of odd perturbations
\be
V_s = \frac{f(r)}{r^2}\left((1-s^2)\left(2\k^2r + (1+\k^2r_+^2)\frac{r_+}{r}\right)+ \n(\n+1) + 3s(s-1)\k^2r^2\right)
\ee
could still be negative just for $s=+2$ (the effective potential for even tensor perturbations \rf{V-2} is positive for any $\n_{lm}\ge 2$). The~condition of $V_{+2}(r)$ negativity is
\be
\n(\n+1)<(1+\k^2r_+^2)\frac{3r_+}{r},
\ee
and $V_{+2}(r)<0$ within $r\in[r_+,r_0)$, where $r_0 = \frac{3r_+(1+\k^2r^2_+)}{\n(\n+1)}$ ($r_0>r_+$).
Therefore, for~eigenvalues $\n$, we get
\be
\n(\n+1)<3(1+\k^2r_+^2).
\la{condAdS}
\ee

Apparently, the~result strongly depends on the specific value of $\k^2r_+^2$ ($r_+$ is the radius of the event horizon), so now $\n$ depends on the size of a BH. In~general, the~value of $\n_{cr}$ in AdS becomes higher than that of Minkowski spacetime; hence, the QNMs with a positive imaginary part may appear for smaller values of $\a$.

To compute the tensor QNMs, here, we will follow the approach of ref.~\cite{Horowitz:1999jd}. In addition, we will take into account the following features of the fundamental (i.e., least damped) QNMs of a Schwarzschild-AdS BH background, marked in ref.~\cite{Cardoso:2003cj}:
\begin{itemize}
	\item for large (with $r_+\kappa \simeq 100$) and intermediate (with $r_+\kappa \simeq 1$) BHs the fundamental quasinormal modes are purely imaginary and scale as $\omega/\kappa \simeq (r_+\kappa)^{-1}$;  
	\item for small BHs the fundamental frequencies get non-trivial real and imaginary parts; the latter behaves as $\text{Im}\,\w\simeq-r_+$. In~the limit $r_+\rightarrow0$, the QNMs turn into normal modes of AdS spacetime, determined by $\w^{AdS}_l = 2n + l + 2$.
\end{itemize}

In the background of a quasi-spherical axisymmetric neutral BH, we have three different tensor QNMs---of frequencies $\w_{20}$, $\w_{21}$ and $\w_{22}$---one of which, $\w_{20}$, coincides with that of the standard spherically-symmetric AdS-Schwarzschild BH. The~other two, $\w_{21}$ and $\w_{22}$, become functions of the deformation degree $\a$. It turns out that, similar to the spherically-symmetric case, frequencies of the fundamental QNMs for large and intermediate BHs come to be purely imaginary. Their functional dependences on $\a$ are plotted in Figure~\ref{fig8}. Looking at the left panel of Figure~\ref{fig8}, one may notice that, for a chosen deformation degree $\a$, the least damped QNM corresponds to $m = 2$ that gives the smallest value of $\nu$. Values of the deformation degree $\a = 0.2$ and $\a = 0.5$ still give the negative imaginary part of the QNM frequencies. However, increasing the value of $\a$ may trigger flipping the sign of $\text{Im}\,\w$. Indeed, in~the right panel of Figure~\ref{fig8}, one finds dependences of $\text{Im}\,\w_{21}$ and $\text{Im}\,\w_{22}$ for a large BH (with $\k r_+ = 10.0$) on the parameter $\a$. Starting from $\a > 1$, $\text{Im}\,\w_{22} $ turns out to be positive. It corresponds to $\nu_{22} < 1$. In~contrast, $\text{Im}\,\w_{21}$ remains negative, even for those $\a$ for which $V_{+2}(r)<0$.

\begin{figure}[ht!]
	\includegraphics[scale=0.08]{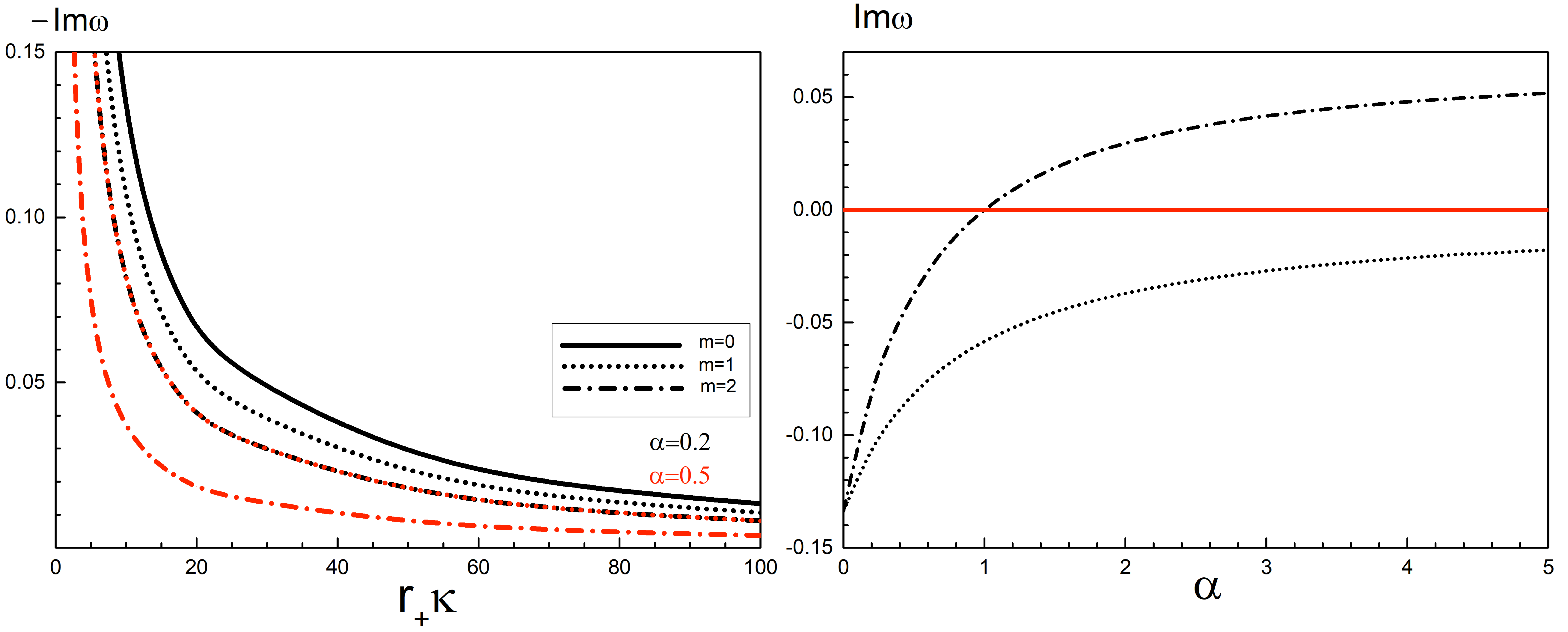}
	\caption{\label{fig8} \textbf{Left panel}: imaginary parts of
	$\w_{2m}$ as functions of $r_+\k$. 
	\textbf{Right panel}: imaginary parts of \\Quasi\-nor\-mal modes (QNMs) of frequencies $\w_{2m}$ ($m = 1, 2$) as functions of the deformation degree $\a$.} 
\end{figure}
\begin{figure}[ht!]
	\includegraphics[scale=0.08]{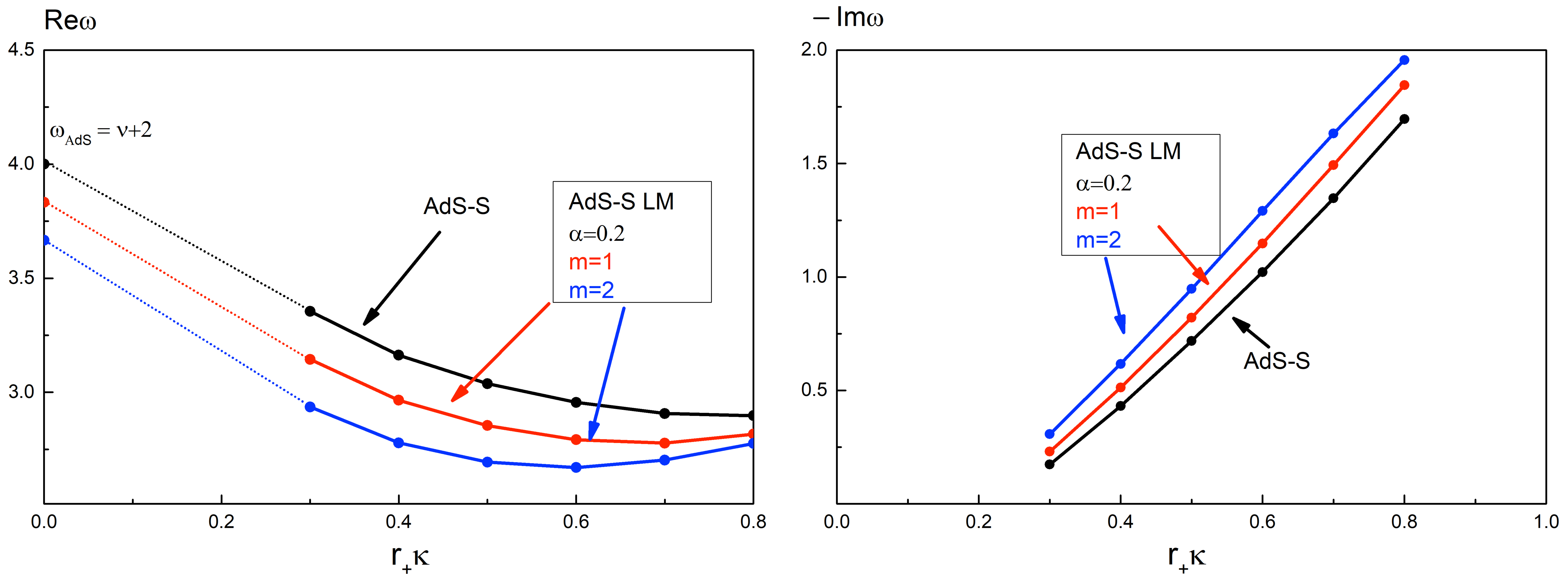}
	\caption{\label{fig9} Functional dependences on
	$r_+\k$ of real (\textbf{left panel}) and imaginary (\textbf{right panel}) parts of the fundamental QNMs of frequencies $\w_{2m}$ ($m = 1, 2$). (Small BHs; the deformation degree is fixed to be $\a = 0.2$.) AdS-S stands for Anti-de Sitter-Schwarzschild; AdS-S LM is the shorthand notation of Anti-de Sitter-Schwarzschild with the Liouville Mode (distorted BH).  } 
\end{figure}

In addition, in~the left panel of Figure~\ref{fig8}, we can find that pure imaginary QNMs of large and intermediate BHs depend on $r_+\k$ quite similarly. This observation allows us to conclude that, for any fixed value of $\a$, one can find the spherically symmetric counterpart (of large and intermediate quasi-spherical BHs) with a larger value of the horizon location $r_+$, which possesses the same fundamental QNMs. The~exception is the $\w_{22}$ mode, the~imaginary part of which becomes positive for $\a >1$, so such a correspondence does not take~place.

Examining the QNMs of small AdS-Schwarzschild BHs, we find they behave much like the standard spherically-symmetric case. One observes that $\text{Im}\,\w\simeq -r_+$ within the range $r_+\kappa\in[0.3; 0,8]$ (right panel of Figure~\ref{fig9}). Once $r_+\rightarrow 0$, the~real parts of the fundamental QNMs are determined by the expression for normal fundamental modes of empty AdS spacetime, $\w_{AdS} = l + 2$, in~which $l$ is replaced with $\n$ (left panel of Figure~\ref{fig9}).

Therefore, as~in the case of flat spacetime, we have established the critical value of the deformation degree $\a=1$, below~which the background geometry of distorted/quasi-spherical BHs is stable against small tensor perturbations, but~above which the background geometry of large quasi-spherical BHs becomes~unstable.

\section{Summary and Open~Questions}\label{sec:7}

Let us make a brief sum up of our achievements and findings.
We have discussed the scattering problem for small perturbations---scalar, vector, and tensor---on quasi-spherical BHs in Minkowski and AdS spacetimes. Studying the problem, in general, has resulted in the following~observations:
\begin{itemize}
\item
There is a deep connection between distorted and quasi-spherical static BHs. To~be precise, we have established the generalization of the Weyl-Erez-Rosen solution to the flat-space Einstein vacuum equations that is reduced to the quasi-spherical Schwarzschild BH solution, the~metric potential of which obeys the Liouville equation.
\item
The obtained BH spacetime is of type D in the Petrov classification. It makes it possible, despite the spherical symmetry breaking, to~separate the variables in dynamical equations of small perturbations and to arrive at the generalized Regge-Wheeler-Zerilli equations, as well as to the generalization of the spherical harmonics equation.
\item
The outcomes of the spherical symmetry breaking are: non-integer eigenvalues in the generalized spectral problem, coming from the angular part of dynamical equations of small perturbations; their multi-dimensional character for each scattering mode, when the dimension of the appropriate set of eigenvalues is determined by the degree of the corresponding spherical harmonics; and last, the~functional dependence of the generalized eigenvalues on the deformation degree parameters.
\end{itemize}

Restoring a part of the spherical symmetry---the axial symmetry of the spacetime background---relevant for a bunch of astrophysical problems has led us to the following~findings:
\begin{itemize}
\item 
The angular dependence of corresponding quantities is further reduced to the dependence on the single polar angle; the spherical Liouville equation for the metric potential turns into the enlargement of general Legendre equation, which can be explicitly solved.
\item
It turns out that the eigenvalues of the generalized spectral problem depend solely on one parameter of the deformation degree after all. And~this functional dependence has been recovered in analytical form. In addition, it has been observed that the generalized eigenvalues are quantized in non-integers.
\item
For every scattering mode corresponding to the appropriate degree of spherical harmonics $l$, there are $l+1$ different values of the grey-body factors, properties of which are determined by the deformation degree. For~instance, we have observed that the grey-body factors increase with increasing the value of the deformation.
\item
Studying the issue of stability of BH backgrounds, we have found that the value of the deformation degree equal to one is the critical value for the stability of the axially symmetric quasi-spherical Schwarzschild BH in Minkowski and AdS spacetimes against the specific small tensor perturbations.
\item
We also find that, for large and intermediate AdS-Schwarzschild quasi-spherical BHs, the fundamental tensor QNMs of any fixed admissible value of the deformation degree are the same as that of a spherically-symmetric BH with a larger value of the event horizon along the radial direction.
\end{itemize}

Therefore, we have observed significant differences in scattering characteristics of gravitational waves caused by losing the spherical symmetry of the background spacetime of their propagation. It would be interesting to find signs of the established effects in the data of real astrophysical observations, at~least for slowly rotating~systems.

Finally, we will touch upon the following point. It is well known the tidal deformation of compact objects in double neutron star systems is described by the famous I-Love-Q relation; see refs.~\cite{Yagi:2013awa,Yagi:2016ejg} for reviews. It seems to be important to figure out any possible relation between the effective metric used in refs.~\cite{Yagi:2013awa,Yagi:2016ejg} and that of the distorted Kerr. Another direction of further studies is related to recovering the metric of a quasi-spherical rotating BH and studying the scattering processes in a more realistic setup. Our preliminary investigations showed the fail of the Newman-Janis algorithm~\cite{Newman:1965tw} (also see refs.~\cite{Drake:1998gf,Ferraro:2013oua,Erbin:2014aya,Nawarajan:2016acj,Erbin:2016lzq} and refs. therein) upon constructing the rotating extension of a quasi-spherical static metric, mostly used in the paper. Perhaps, the~observed here connection between distorted and quasi-spherical BHs will make possible to complete this task in another way. We hope to continue studies on this and other related topics and report results in forthcoming~publications.

\bsk
{\bf Acknowledgments}: A.M.A. is grateful to the Institute for Theoretical Physics of NSC KIPT, where part of this work was completed, for its warm hospitality. A.J.N. acknowledges all the colleagues, with whom the subject of the present research was discussed, for fruitful and stimulating conversations. The authors are thankful to the anonymous reviewers for their suggestions in improving the early version of the paper.


\bsk

\appendix

\renewcommand{\theequation}{A.\arabic{equation}}
\setcounter{equation}{0}

\section*{Appendix A. Spherically symmetric and axially symmetric solutions to the Liouville equation}\label{app1}

First, let us briefly discuss the way to solve the spherical Liouville Equation \rf{Lspheq} in terms of unconstrained functions. See Appendix B of ref.~\cite{Moskalets:2016uno} for more details.

To this end, it is convenient to turn to the stereographic projection plane coordinates 

\be
z = e^{i\varphi}\tan\frac\theta2, \qquad  \bar{z}= e^{-i\varphi}\tan\frac\theta2 \, .
\la{zbarzdef}
\ee

Then, the angular part of $S^2$ line element, $ds^2=d\th^2+\sin^2\th d\vf^2$, becomes the Fubini-Study metric of $CP^1$ complex projective space:
\be
ds^2 =  \frac{4}{(1 + z\bar{z})^2}\,dz\,d\bar{z},
\la{FS}
\ee	
and the angular part of the Laplacian \rf{Laplac} is simplified up to	
\be	
\Delta_{\theta,\varphi} = \frac14(1 + z\bar{z})^2\partial_z\partial_{\bar{z}}.
\la{S2zz}
\ee

Consequently, the~spherical Liouville Equation \rf{Lspheq} for $\chi(\th,\varphi)$ takes the form of
\be
\frac14(1 + z\bar{z})^2\partial_z\partial_{\bar{z}}\chi(z,\bar{z}) + 2e^{\chi(z,\bar{z})} - 2 = 0
\la{Liouzz}
\ee
and can be solved in terms of arbitrary complex analytical function $F(z)$ \cite{Moskalets:2017koi,Moskalets:2016uno}:
\be
\chi(z,\bar{z}) = -2\ln\left[F(z)\bar{F}(\bar{z})+1\right] + \ln\left[\fr{dF(z)}{dz}\fr{d\bar{F}(\bar{z})}{d{\bar z}}\right]+ \ln(1+z\bar{z})^2. 
\label{chisolzz}
\ee

Second, let us survey the way of getting expression \rf{expaxs} in the case of axial symmetry of a quasi-spherical BH. Turning to the isothermal coordinates $(x, y)$ in $z$, viz. $z = x + iy$, for~\rf{Liouzz}, we obtain
\be
\frac14(1 + x^2 + y^2)^2\left(\frac{\partial^2\chi}{\partial x^2} + \frac{\partial^2\chi}{\partial y^2}\right) + 2e^{\chi(x,y)} - 2 = 0\,.
\label{chiEqxy}
\ee

The Fubini-Study metric \rf{S2zz} comes to
\be
ds^2 = e^{\chi_0(x,y)}(dx^2 + dy^2),
\ee
with
\be
e^{\chi_0(x,y)} \equiv \frac{4}{(1 + x^2 + y^2)^2}, 
\ee
or
\be
\chi_0(x,y) = 2\ln\left[\frac{2}{1+ x^2 + y^2}\right].
\ee

Next, we define a function $\Phi(x,y)$ as 
\be
\chi(x,y) = \Phi(x,y) - \chi_0(x,y) = \Phi(x,y) - 2\ln\left[\frac{2}{1+ x^2 + y^2}\right].
\la{chixydefzz}
\ee

Substituting \rf{chixydefzz} into (\ref{chiEqxy}) leads to the following equation for $\Phi(x,y)$:
\be
\frac14\bigg(1 + x^2 + y^2\bigg)^2\left[\frac{\partial^2\Phi}{\partial x^2} + \frac{\partial^2\Phi}{\partial y^2} + 2e^{\Phi(x,y)}\right] = 0 \quad \leadsto \quad  \frac{\partial^2\Phi}{\partial x^2} + \frac{\partial^2\Phi}{\partial y^2} + 2e^{\Phi(x,y)} = 0.
\ee

Now, the~axially-symmetric case corresponds to demanding $\Phi(x,y)$ to solely depend on the radial coordinate $\rho$, related to $(x,y)$ via $\rho^2 = x^2 + y^2$. Hence, $\Phi(x,y)  \ra \Phi(\rho)$, and~$\Phi(\r)$~obeys 
\be
\frac{d^2\Phi(\rho)}{d\rho^2} + \frac{1}{\rho}\frac{d\Phi(\rho)}{d\rho} + 2e^{\Phi(\rho)} = 0.
\la{Phiaxeq}
\ee

With $\Phi = \ln f(\rho)$, Equation \rf{Phiaxeq} transforms into
\be
f\fr{d^2f(\rho)}{d\r^2} - \left(\fr{df(\r)}{d\r}\right)^2 + \frac{f}{\rho}\,\fr{df(\rho)}{\r} + 2f^3(\r) = 0,
\ee
the solution to which is
\be
f(\rho) = \frac{2+C_1}{2\rho^2\cosh^2\left(\frac{\sqrt{2+C_1}(C_2 -\ln\rho)}{\sqrt{2}}\right)} .
\ee

Turning to Equation \rf{chixydefzz} back, for~$\chi(\r)$, we have
\be
\chi(\rho) = \ln\left[\frac{(2+C_1)(1 + \rho^2)^2}{8\rho^2\cosh^2\left(\frac{\sqrt{2+C_1}(C_2 -\ln\rho)}{\sqrt{2}}\right)}\right].
\ee

Finally, with~inserting new constants $a = \sqrt{1 + \frac{C_1}{2}}$, $b = e^{aC_2}$ and replacing $\r$ with its functional dependence on $\th$, $\rho = \tan\frac\th2$ (cf. \rf{zbarzdef}), we arrive at
\be
e^{\chi(\th)} = \left(\frac{a}{b}\right)^2\tan^{2a-2}\frac\th2\left(\frac{1+\tan^2\frac\th2}{(1+b^{-2}\tan^{2a}\frac\th2)}\right)^2 .
\la{expchithsol}
\ee 

On account of the trigonometric identity $\tan^2 \th/2=(1-x)/(1+x)$, $x=\cos \th$,
the obtained solution \rf{expchithsol} to the Liouville equation in polar coordinates turns into \mbox{expression~\rf{expaxs}}.

\end{document}